\documentclass{article}

\usepackage{arxiv}

\usepackage[utf8]{inputenc} % allow utf-8 input
\usepackage[T1]{fontenc}    % use 8-bit T1 fonts
\usepackage{hyperref}       % hyperlinks
\usepackage{url}            % simple URL typesetting
\usepackage{booktabs}       % professional-quality tables
\usepackage{amsfonts}       % blackboard math symbols
\usepackage{nicefrac}       % compact symbols for 1/2, etc.
\usepackage{microtype}      % microtypography
\usepackage{graphicx}
\usepackage[numbers]{natbib}
\usepackage{doi}
\usepackage{xcolor}

\usepackage{natbib}
\usepackage{hyperref}
\usepackage{upgreek}
\usepackage{sidecap}
\usepackage{acronym}    % allows consistent definition of acronyms used throughout the text
\acrodef{CD}[CD]{contrast detection}
\acrodef{DVS}[DVS]{dynamic vision sensing}
\acrodef{EBV}[EBV]{event-based vision}
\acrodef{CFD}[CFD]{computational fluid dynamics}
\acrodef{DSR}[DSR]{dynamic spatial range}
\acrodef{DVR}[DVR]{dynamic velocity range}
\acrodef{FWHM}[FWHM]{full-width-half-mean}
\acrodef{HP-LED}[HP-LED]{high-power LED}
\acrodef{HWA}[HWA]{hotwire anemometry}
\acrodef{IPCT}[IPCT]{image pattern correlation technique}
\acrodef{LED}[LED]{light emitting diode}
\acrodef{LES}[LES]{large-eddy simulation}
\acrodef{PIV}[PIV]{particle image velocimetry}
\acrodef{ppp}[ppp]{particles per pixel}
\acrodef{PTV}[PTV]{particle tracking velocimetry}
\acrodef{PWM}[PWM]{pulse width modulation}
\acrodef{rms}[rms]{root mean square}
\acrodef{2D-2C-PIV}[2D-2C-PIV]{two-dimensional (planar), two component particle image velocimetry}
\acrodef{ROI}[ROI]{region of interest}
\begin{document}

%%\unnumbered% uncomment this for unnumbered level heads

\title{Event-based imaging velocimetry using pulsed illumination}

\date{November 10, 2022}	% Here you can change the date presented in the paper title
%\date{} 					% Or removing it

\author{ \href{https://orcid.org/0000-0002-1668-0181}{\includegraphics[scale=0.06]{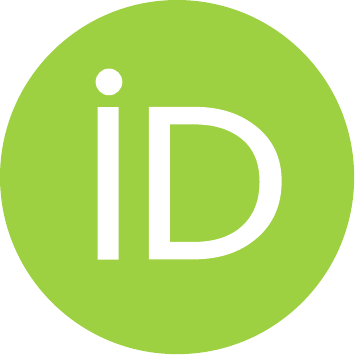}\hspace{1mm}Christian E. Willert} \\
	Engine Measurement Techniques\\
	DLR Institute of Propulsion Technology\\
	German Aerospace Center\\
	51170 K\"{o}ln, Germany \\
	\texttt{chris.willert@dlr.de} 
}

% Uncomment to override  the `A preprint' in the header
\renewcommand{\headeright}{ }
\renewcommand{\undertitle}{under consideration for publication in \emph{Experiments in Fluids}}
\renewcommand{\shorttitle}{EBIV -- Event-based Imaging Velocimetry}

%% my stuff...
\newcommand{\hlt}[1]{\textcolor{blue}{#1}} % to turn it off
\newcommand{\hltred}[1]{\textcolor{red}{#1}}
\newcommand{\todo}[1]{\textcolor{red}{\textbf{To do: }#1}}

%%==================================%%
%% sample for unstructured abstract %%
%%==================================%%
\hypersetup{
	pdftitle={Event-based imaging velocimetry using pulsed illumination},
	pdfsubject={},
	pdfauthor={Christian Willert},
	pdfkeywords={fluid flow measurement, particle imaging, event-based vision sensing, dynamic vision sensor, neuromorphic imaging, high-speed imaging, time-resolved PIV, particle tracking, PTV},
}

\maketitle

\begin{abstract}
The paper addresses the shortcoming of current event-based vision (EBV) sensors in the context of particle imaging.
Latency is introduced both on the pixel level as well as during read-out from the array and results in systemic timing errors when processing the recorded event data.
Using pulsed illumination, the overall latency can be quantified and indicates an upper bound on the frequency response on the order of 10--20~kHz for the specific EBV sensor.
In particle-based flow measurement applications, particles scattering the light from a pulsed light source operating below this upper frequency can be reliably tracked in time.
Through the combination of event-based vision and pulsed illumination, flow field measurements are demonstrated at light pulsing rates up to 10~kHz in both water and air flows by providing turbulence statistics and velocity spectra.
The described EBV-based velocimetry system consists of only an EBV camera and a (low-cost) laser that can be directly modulated by the camera, making the system compact, portable and cost effective.
\end{abstract}

%\section*{Graphical abstract}
\centerline{
\includegraphics[width=0.9\columnwidth]{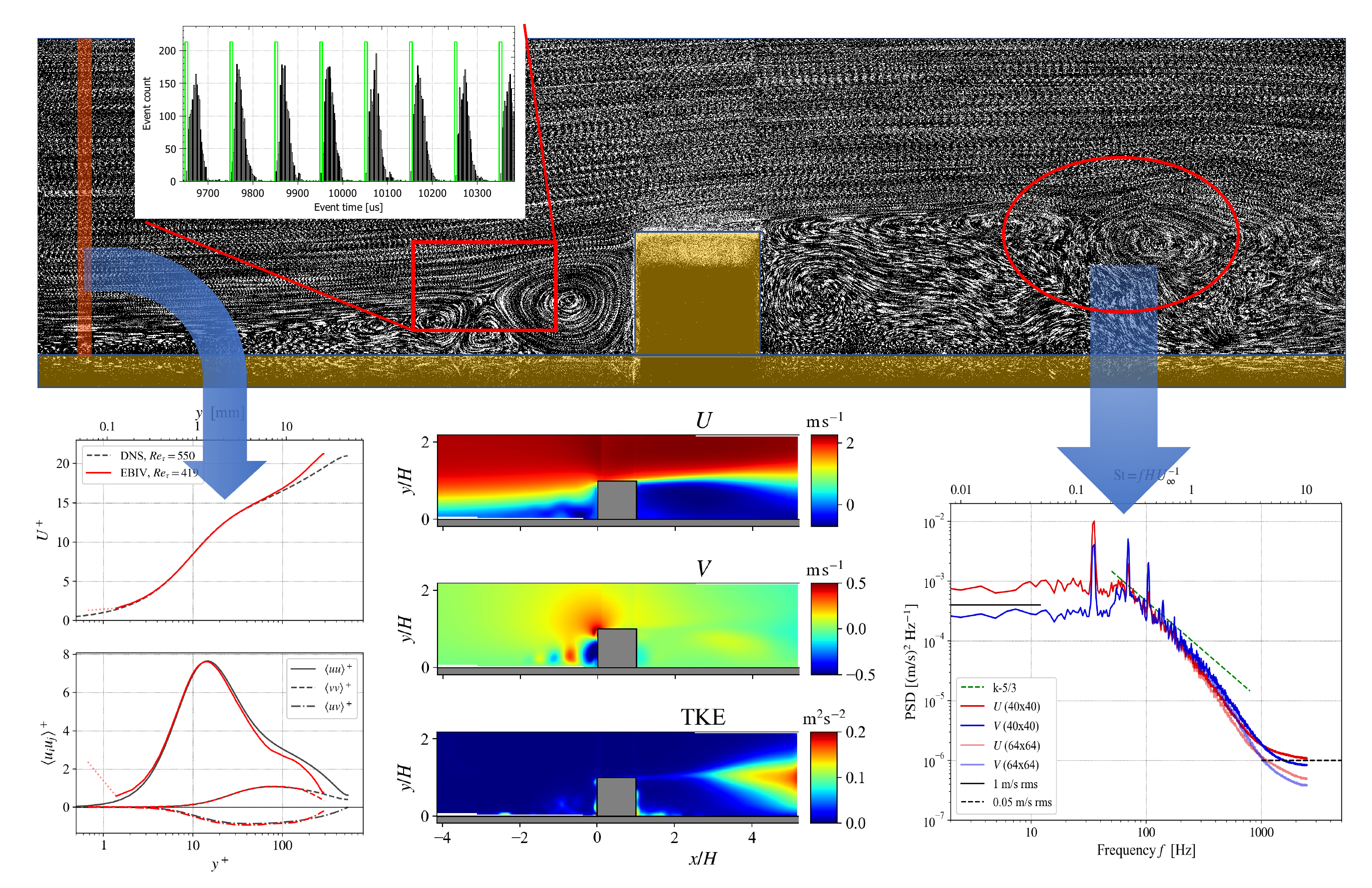}
}

% keywords can be removed
\keywords{event-based vision \and dynamic vision sensor  \and latency  \and
	 particle imaging  \and PIV  \and particle tracking velocimetry  \and PTV  \and 
	 pulsed illumination  \and structured light}

%%\pacs[JEL Classification]{D8, H51}

%%\pacs[MSC Classification]{35A01, 65L10, 65L12, 65L20, 65L70}

\section{Introduction}
\label{Sec:Intro}

\Ac{EBV}, also referred to as \ac{DVS} or neuromorphic imaging, constitutes a new approach to motion-related imaging.
Unlike conventional frame-based image sensors, an \ac{EBV} sensor only records intensity changes, that is, contrast change events, on the pixel level.
As each pixel operates on its own, the resulting event data stream is asynchronous in nature, varying from quiescent in static imagery and increasing to considerable data rates for scenes with high dynamic content.
Consequently, the associated event-data processing requires a partial departure from the established concepts of frame-based image processing.

The underlying concepts were originally proposed by Carver Mead and Misha Mahowald \citep{Mahowald:1992} in the early 1990's and were aimed at mimicking the functionality of the eye's retina - hence coining the name ``silicon retina" in their work.
By the late 2000's the technology had matured enough to result in usable prototype cameras based on event-based sensors with pixel area on the order of 128$\times$128 elements \citep{Lichtsteiner:2008}.
More recently, several companies have introduced fully integrated commercial grade event-cameras with detector arrays up to 1~MPixel \citep{Finateu:2020}.
This improved accessibility to event cameras has greatly expanded the areas of event-based vision applications and has made the present study possible.
Comprehensive overviews of the current status of event-based vision, its range of applications and associated data processing  are provided by \cite{EBVreview:2022} as well as \cite{Tayarani-Najaran:2021}.
Beyond this, a Github repository maintained by the Robotics and Perception Group of the University of Zurich provides an extensive collection of literature and resources on the subject \citep{GIT:EBVrefs}.

While the number of applications for \ac{EBV}/\ac{DVS} are continually increasing, the latency between the \ac{CD} event and the actual time assigned to it by the detector hardware has been identified as one of the more important shortcomings \citep{Bouvier:2021}.
Data sheets and literature quote latency values on the order of 3 to 200~$\upmu$s (see Table~1 in \citealp{EBVreview:2022}).
In practice, this value may even be higher and depends on lighting conditions, various adjustable detector parameters (\textit{biases}) as well as the number of events per unit time.
As a consequence any sort of data processing relying on the actual event-time will be affected by temporal uncertainty.
In the case of particle-based fluid flow measurement the particle velocity estimate will have an uncertainty that is directly related to the uncertainty of the event time-stamps.

In the following, the working principles of event-based imaging are briefly outlined before presenting measurements of the latency of the currently used hardware.
The second part describes the use of pulsed illumination to obtain more precise information on the particle-produced events.
The pulsed-illumination event-based imaging concept (\emph{pulsed-EBIV}) is demonstrated on simple water and air flows containing a wide range of flow dynamics.
Along with comparative PIV measurements, the statistical and spectral analysis of the processed EBIV data allows an assessment of the overall measurement uncertainty of the proposed technique.
The material presented herein extends on a recent publication on the subject \citep{Willert_EBIV:2022}. % and is intended to extend upon the previous publication.

\section{Event sensor operation and signal processing}
\label{sec:ebi-pixel}
\begin{figure}[htb]
\centering
\includegraphics[width=0.99\columnwidth]{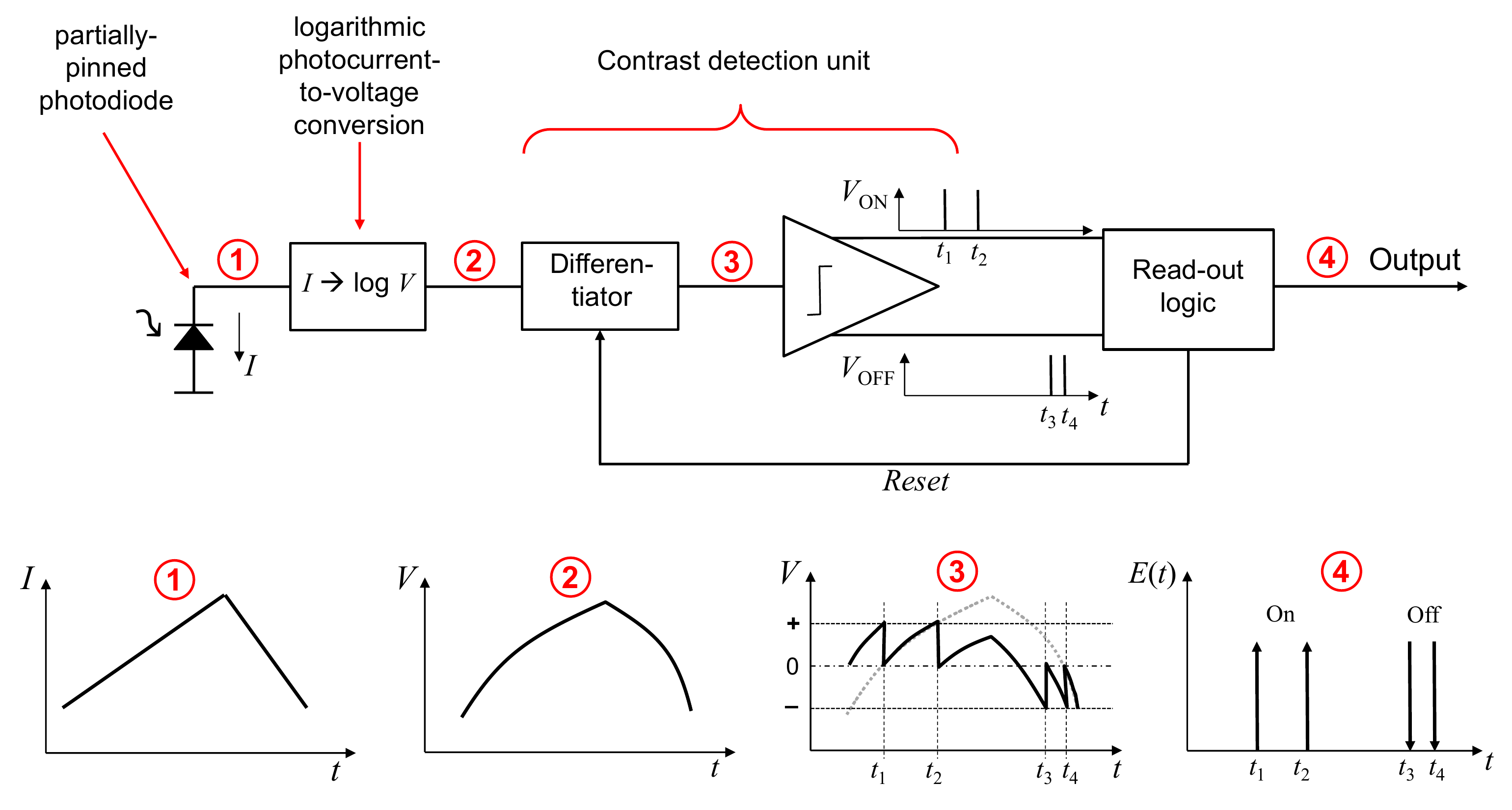}
\caption{Simplified schematic of an EBV pixel and associated signals at each step.}
    \label{fig:ebv-pixel}
\end{figure}

Each \ac{EBV} pixel consists of a photodiode with adjoining contrast change detection unit and logical unit (Fig.~\ref{fig:ebv-pixel}).
Photons collected by the photodiode induce a photocurrent that is converted to a voltage.
The logarithm of this voltage is supplied to the contrast detection unit allowing it to trigger \acf{CD} events at a fixed relative contrast threshold, typically on the order of 10-25\%.
A comparator unit distinguishes between positive and negative intensity change events, where \emph{positive} refers to an intensity change from dark to bright (``On" event) whereas \emph{negative} indicates a reduction of intensity (``Off" event).
Once a pixel detects a positive or negative event the logical unit sends out an acknowledge request to the detector's event monitors, termed \emph{arbiters}, which then register the event and transmit the event's time-stamp $t_i$, its polarity $p_i$ and pixel coordinates $(x_i,y_i)$ to the output data stream.
After a pending event has been acknowledged, the logical unit resets the detector part of the pixel allowing a further event to be captured.

\begin{SCfigure}[\sidecaptionrelwidth][htb]\centering
\includegraphics[width=0.6\columnwidth]{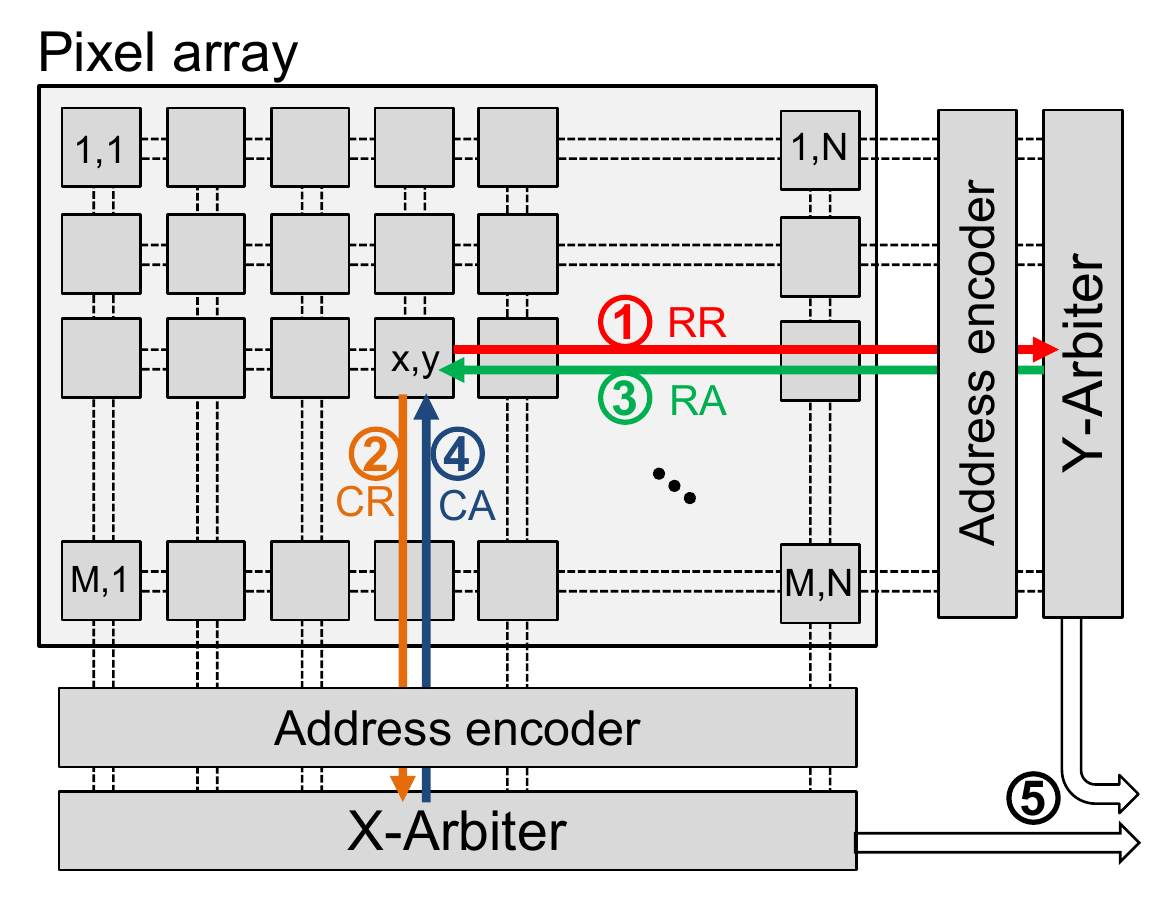}
\caption{Simplified schematic asynchronous readout system of an EBV sensor (following Fig.II.12 from \citealp{Bouvier:2021}).
An event-triggered pixel at $(x,y)$ sends row and column requests (RR, CR) that are acknowldged (RA,CA) by the arbiters.}
    \label{fig:ebv-sensor-readout}
\end{SCfigure}

The asynchronous readout procedure is outlined in Fig.~\ref{fig:ebv-sensor-readout}: an event-triggered pixel at position $(x,y)$ sends out both a row request (RR) and column request (CR). 
The requests are  acknowledged (RA,CA) in succession by the respective row and column arbiters. Once the event is fully acknowledged its coordinate, polarity and time stamp is forwarded to the output stream. 
Actual implementations differ between various devices. For instance, the Gen4 device by Prophesee only uses a single arbiter, assigning pending events from the same row with the same time stamp \citep{Finateu:2020}.
The ``first-come-first-served" concept of current event-sensor readout architectures is one of the primary sources of the system-inherent latency.

The sensitivity of the detector array in triggering positive or negative \ac{CD} events is controlled through a number of \emph{biases}. These can be tuned by the user to suit a given application.
The most relevant biases for the present application were found to be the positive and negative detection thresholds along with the refractory period that controls the dead time of a given pixel after registering a \ac{CD} event.
A systematic study of the influence of the biases on event generation and the associated latency is challenging in itself, not only because of the large parameter space, but also because the lighting conditions along with motion of the imaged object(s) contribute.
A detailed overview of event generation and measurement schemes to determine the latency are provided by  \cite{Joubert:2019}.

\subsection{Limitations of EBV sensors}
Within the context of accurate event capture, the latency, that is, the time between the \ac{CD} appearing on the pixel and its actual acknowledgement by the event processing unit, is the most relevant.
The total latency is a combination of the pixel latency and system latency, the first of which is the delay of the pixel itself in responding to a contrast change. This delay differs between pixels and results in a temporal jitter.

\begin{figure}[htb]
\includegraphics[width=\columnwidth]{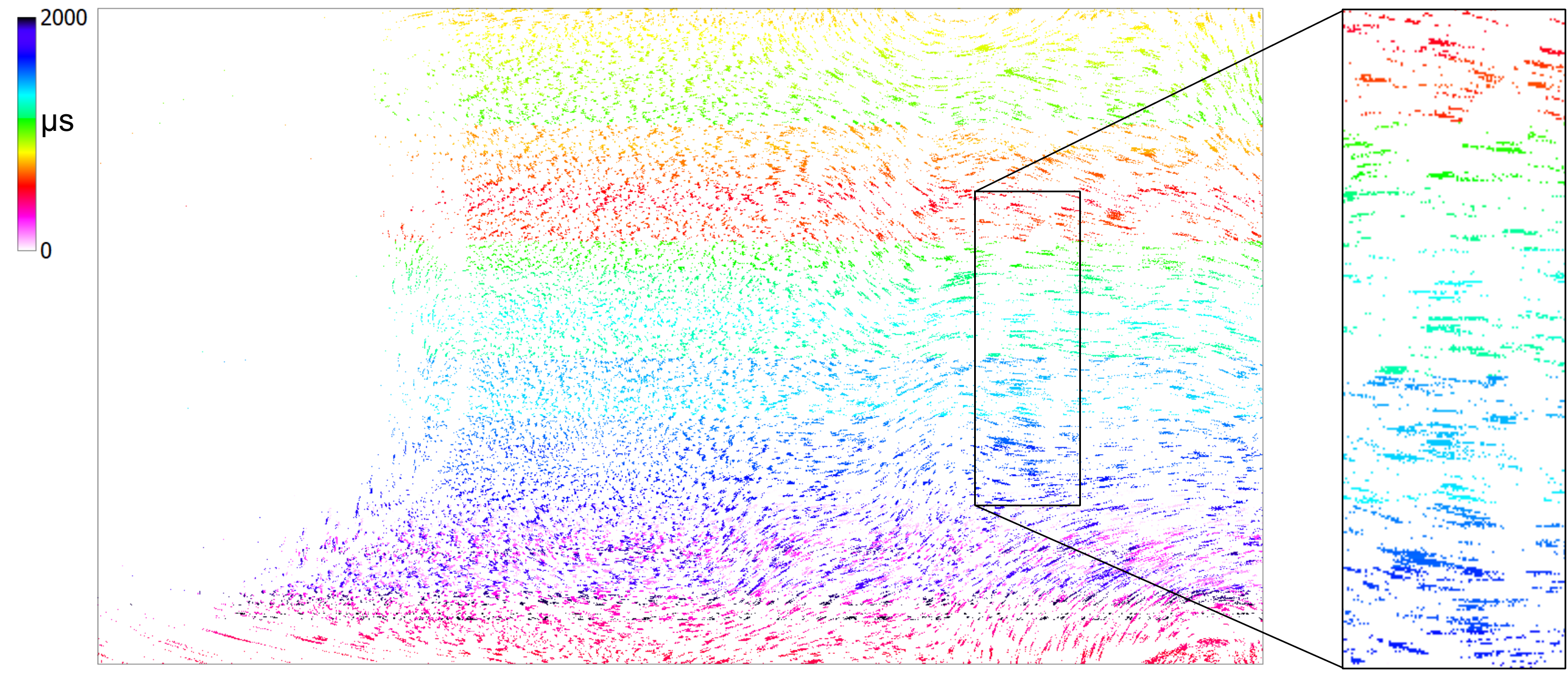}
\caption{Example of the loss of event-timing due to saturation of event-monitoring units (arbiters) within the sensor itself. The pseudo-image contains 2~ms of \ac{CD} events acquired with a continuously operating laser in the wake of a circular cylinder in a small wind tunnel.
The event data rate is about $60\!\cdot\! 10^6$\,events/s.
}
    \label{fig:arbiter-event-overrun}
\end{figure}

With increasing number of \ac{CD} events per unit time, the arbiters begin saturate and can no longer react promptly to pending event requests. This results in a latency between the time assigned to the event and the actual occurrence of the event, and may sometimes reach up to several hundred $\upmu$s.
The event overload may manifest itselfs differently and depends on the sensor architecture and is one of the primary challenges in low-latency event-sensor design.
For the presently used hardware, the arbiter saturation results in horizontal stripes of \ac{CD} events all of which are assigned with the same event time (see Fig.~\ref{fig:arbiter-event-overrun}).
The effect can be monitored with histograms of the event arrival times.
The event rate can be fine-tuned not to result in arbiter overloading by changing the light intensity -- here, laser power and/or lens aperture -- or by adjusting the particle seeding density.
For the presently used Gen~4 HD sensor (Prophesee EVK2-HD,  $1280 \!\times\! 720 $ pixels) up to 50--60 ev/(pixel $\cdot$ s) were found to produce useful particle event data.
Alternatively, a reduction of the \ac{ROI} on the sensor, that is, the number of pixels being monitored, can be applied to increase the event-count per pixel as the arbiters can process a reduced number of pixels more efficiently.

%\begin{figure}[htb]
%\small
%\textbf{a)}\hspace{0.49\columnwidth}\textbf{b)}\\
%\includegraphics[width=\columnwidth]{figs/roi_event_recording}
%\caption{Reduction of the number of active event pixels allows to locally increase the event-count per pixel, \textbf{a:} ROI reduction in the vertical direction, \textbf{b:} reduction to a narrow vertical strip.}
%    \label{fig:roi-event-recording}
%\end{figure}

% \includegraphics[trim=left bottom right top, clip]{file}
%\begin{figure}[htb]
%\begin{SCfigure}[\sidecaptionrelwidth][htb]
%\small
%\includegraphics[trim=50 50 100 50,clip,width=0.6\columnwidth]{figs/cylwake/cylwake_events_per_px}
%\caption{Comparison of the event rates for different event imaging conditions: \textbf{a:} continuous illumination without as reported in \cite{Willert_EBIV:2022},
%\textbf{b:} pulsed illumination, $f_p = 5$\,kHz.
%\todo{repeat using shorter pulse duration}}
%    \label{fig:event-rate-comparison}
%\end{SCfigure}

\section{Event-based imaging with pulsed illumination}
\label{sec:event-based-imaging}
The combination of event-based imaging with pulsed illumination, also termed ``structured light", is not new and already has found several applications in the area of 3D reconstruction.
Using a pulsed laser line at frequencies in the 100--500\,Hz range, \cite{Brandli:2014} reconstructed 3D surfaces from the events produced by the projected line.
\cite{Huang:2021} used a high speed digital light projector to produce a blinking pattern at 1000\,Hz that is imaged by an event camera to reconstruct 3D surfaces.
Event-based imaging in combination with two triangularly modulated lasers of different wavelength are used for \emph{event-based bispectral photometry} to perform surface reconstruction in water by making use of the wavelength dependent absorption characteristic of the medium \citep{Takatani:2021}.
It this case, modulation frequencies of 1--120\,Hz are applied.

\begin{SCfigure}[\sidecaptionrelwidth][htb]
\small
\includegraphics[width=0.6\columnwidth]{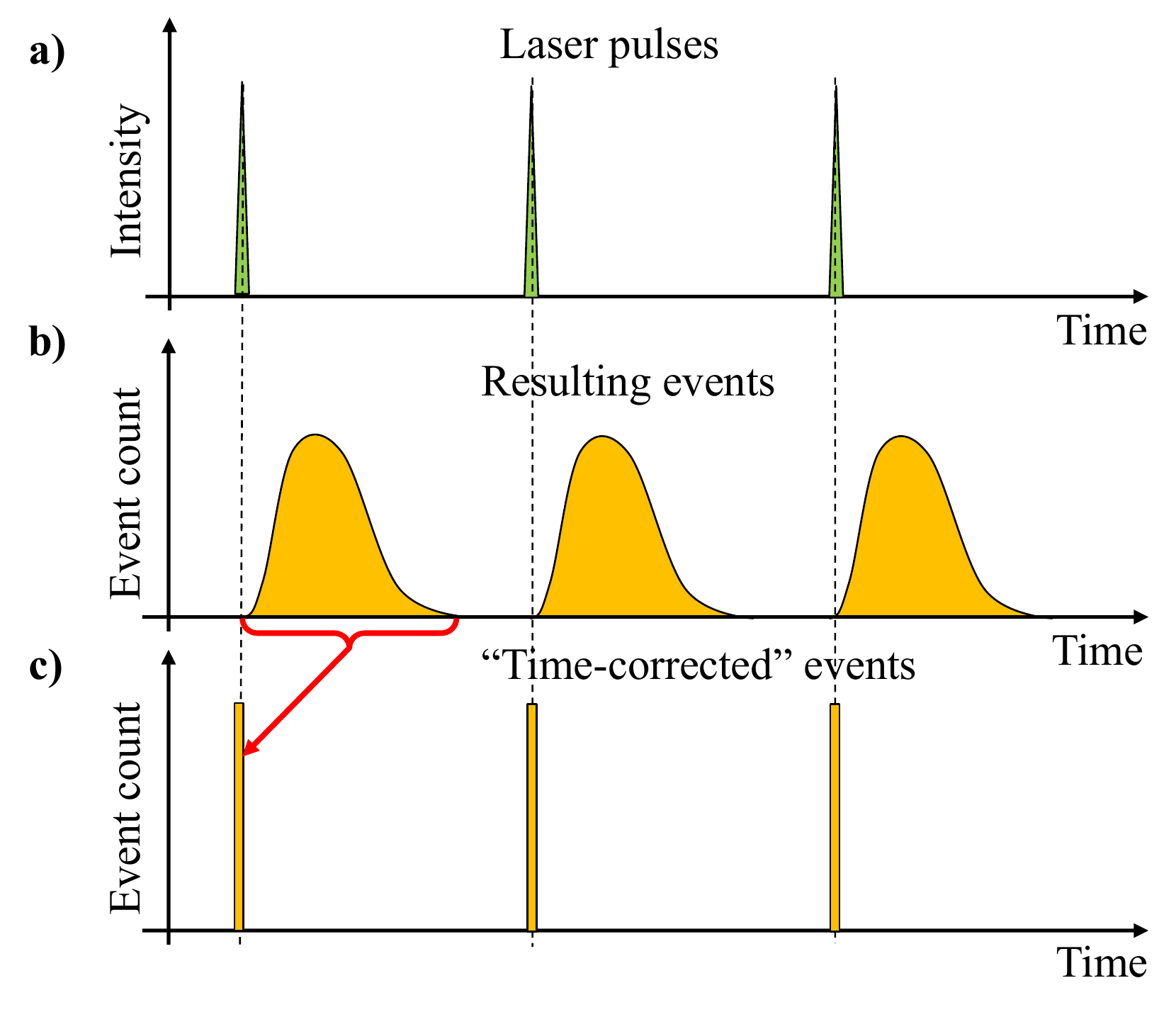}
\caption{Concept of event-based imaging with pulsed illumination: pulses of light (\textbf{a}) generate a burst of events with system specific latency (\textbf{b}). The events generated within a given pulse period are time-stamped with the time of the preceding light pulse (\textbf{c}).}
    \label{fig:pulsed_illumination}
\end{SCfigure}

In the present study, pulsed illumination of a flow seeded with light scattering particles intends to address several issues at the same time:
\begin{itemize}
    \item With continuous illumination stationary or slowly moving particles produce no or very few events. Light pulses will make these particles visible.
    \item The events generated by moving particles decrease with increasing velocity simply because the respective pixel cannot collect a sufficient number of photo-electrons as the particle image passes by. With pulsed illumination the motion of the particles is essentially ``frozen''.
    \item Events generated by pulses of light can be directly associated with the timing of the preceding pulse which reduces the latency issues described in the previous section.
\end{itemize}
The concept of event-based imaging with pulsed illumination is outlined in Fig.~\ref{fig:pulsed_illumination}.
Due to latency the events appear after a light pulse illuminates the scene.
If the latency is shorter than the pulsing interval, then the events can be uniquely associated with the preceding light pulse.
Thereby the uncertainty introduced by the latency can be mitigated.
When the latency exceeds the pulsing interval then events from a given light pulse will appear in the following pulsing interval(s). This overlap will inadvertendly result in timing ambiguities.

\subsection{Event latency measurements}
\label{sec:meas-latency}
\begin{SCfigure}[\sidecaptionrelwidth][htb]
\includegraphics[width=0.6\columnwidth]{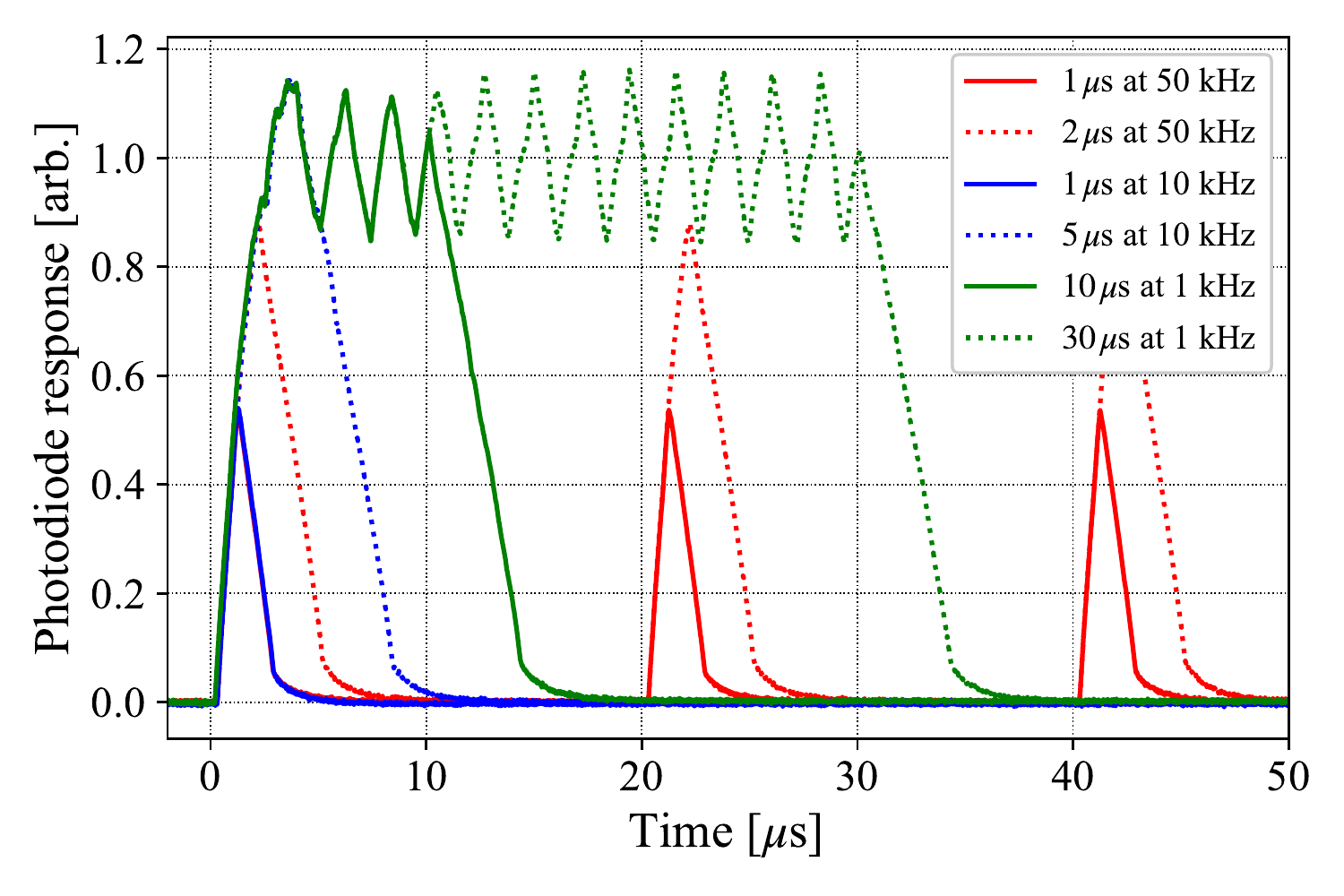}
\caption{Light pulses emitted by a low-cost laser (NEJE E30130) supplied with trigger pulses of different pulse widths and pulse frequencies.}
    \label{fig:laser-pulses}
\end{SCfigure}

The combined latency, consisting of the contribution of both the on-pixel delay and the latency introduced by the arbiters, is estimated by illuminating a particle-seeded fluid flow with a pulsed light source.
The light pulses clearly define time-intervals during which the detector should produce events and should be sufficiently short.
In the present case, a small laser with \ac{PWM} is used to produce light pulses with a duty cycle of 1\% -- 10\% at frequencies of 500\,Hz $ \!< \! f_p \!<\! 20$\,kHz.
The low-cost laser (NEJE Tool E30130) is originally purposed for laser-cutting and engraving. It has a mean output power of 5.2\,W at a wavelength of 450\,nm with a \ac{FWHM} spectral width of 2\,nm as measured with a spectrometer (Ocean Optics USB2000+, 0.5\,nm resolution).
The light pulses emitted by the laser have rise and fall times of about 3\,$\upmu$s (see Fig.~\ref{fig:laser-pulses}).
For pulse widths $\tau_p \!>\! 5\,\upmu$s the light pulses exhibit a saw-tooth modulation of about 15\% of the maximum pulse power with a modulation frequency of about 50\,kHz.
The pulsing frequency has no effect on the shape or amplitude of the emitted laser pulse.
The finite pulse width may have to be taken into consideration as it can introduce particle image streaking on the event detector for fast moving particles.
Here, a Q-switched high-speed solid-state laser with pulses in the sub-microsecond range would improve the measurements.

A water flow experiment (jet in water) is used 
for the assessment of the temporal event generation by pulsed light illumination. The laser light is spread into a parallel light sheet of about 80\,mm height and 1\,mm waist thickness using a $f \!=\! -40$\,mm plano-concave cylindrical lens along with a $f \!=\! 40$\,mm plano-convex cylindrical lens.
A small pump is placed in one corner of a small glass water tank ( $300 \! \times\! 200 \!\times\! 60$ mm$^3$, $H \!\times\! W \!\times\! T$) and produces a turbulent jet of about 1.5\,m/s exit velocity (diameter 8\,mm).
The water is seeded with neutrally buoyant 10\,$\upmu$m silver coated glass spheres.
Event-recordings for different pulse frequencies are acquired near the jet nozzle.
An objective lens with 55\,mm focal length (Nikon Micro-Nikkor 55 mm 1:2.8) with the aperture set at $f_\# \!=\! 5.6$ collects the scattered light at a magnification of 27.0\,pixel/mm ($m \!=\! 0.12$).
The camera event detection thresholds are adjusted to favor positive events.
Camera biases, magnification, lens aperture and seeding concentration are held constant for the measurements.

\begin{figure}[htb]
\small
\textbf{a)}\hspace{0.45\columnwidth}\textbf{b)}\\
\includegraphics[width=0.49\columnwidth]{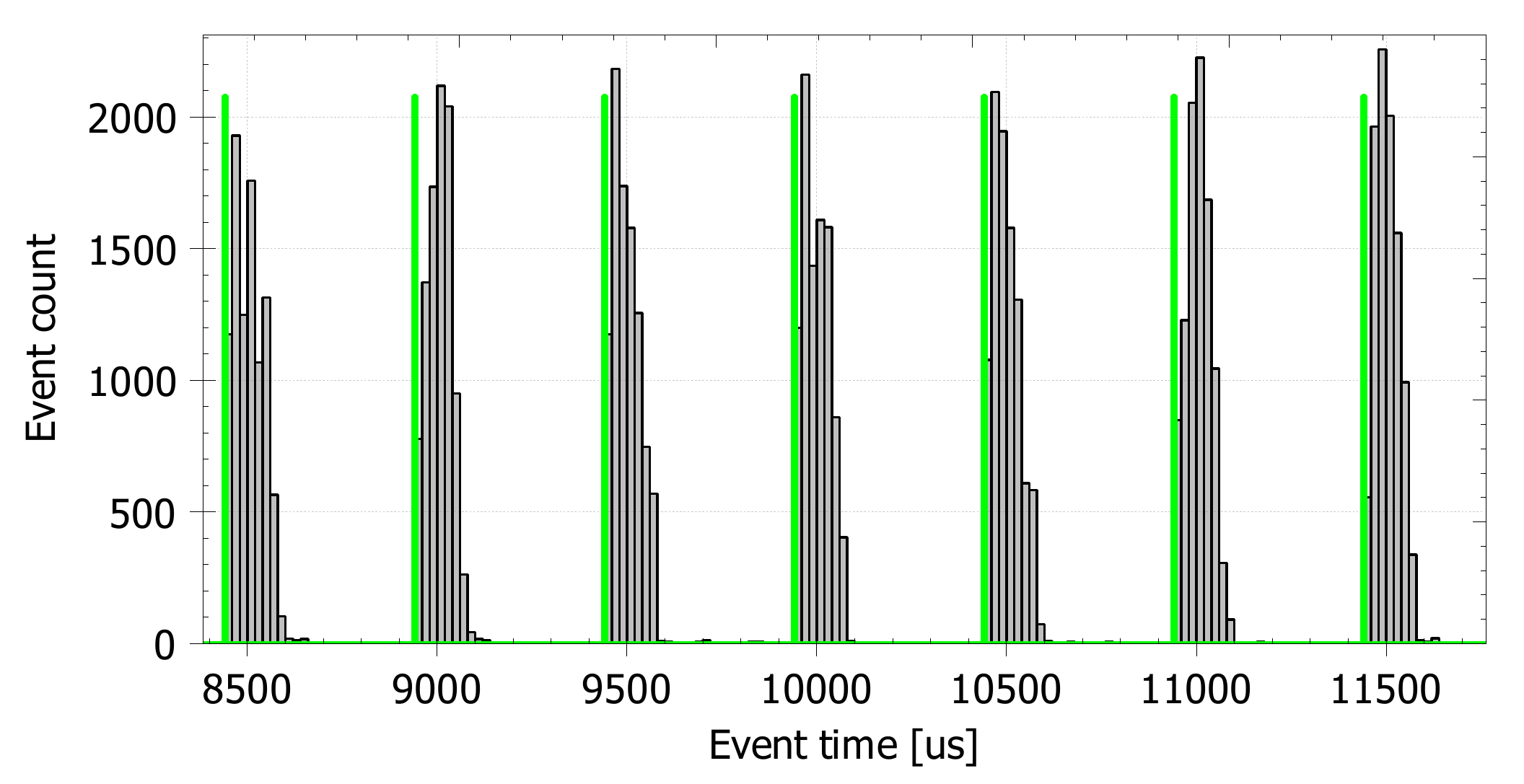}
\includegraphics[width=0.49\columnwidth]{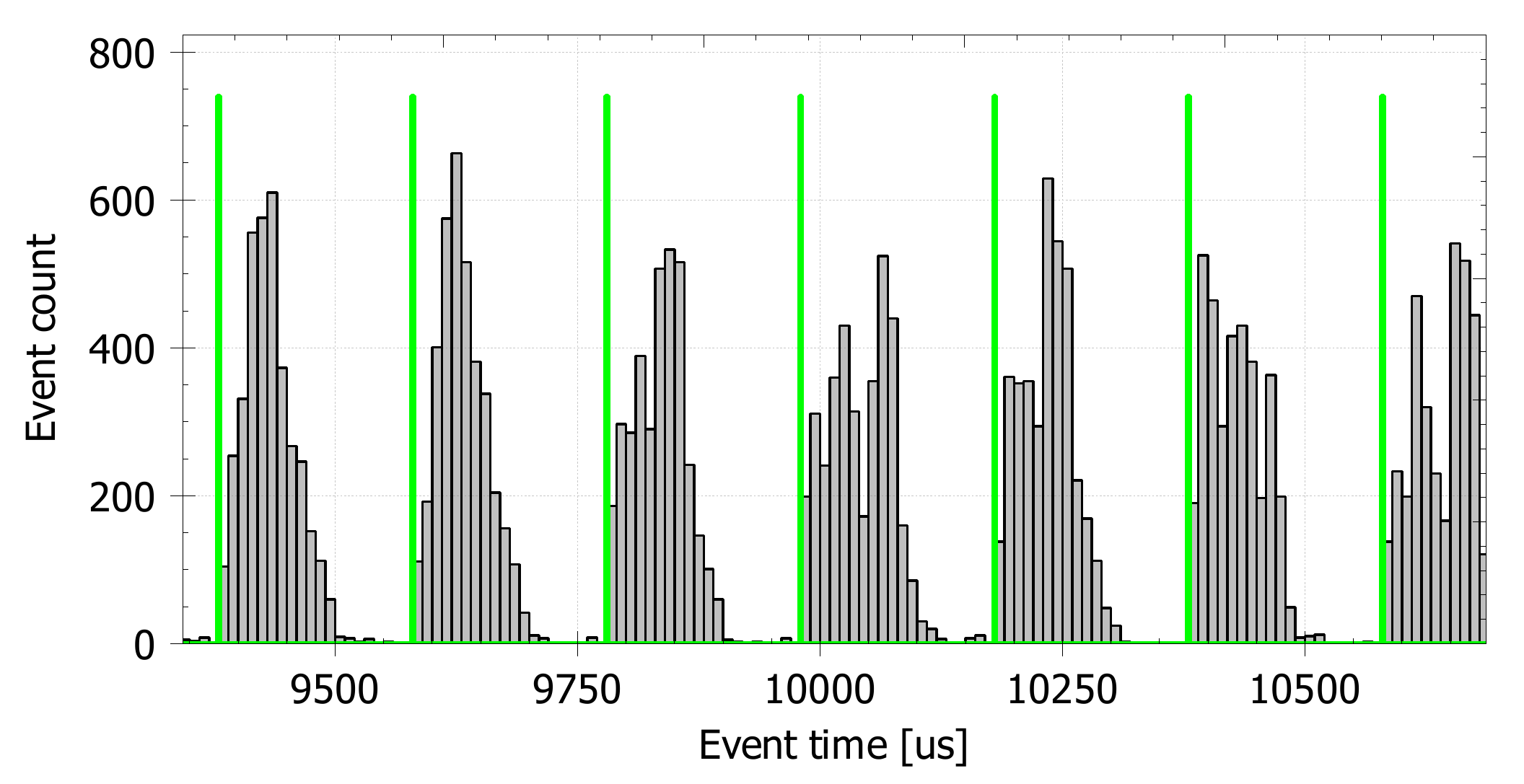}\\
\textbf{c)}\hspace{0.45\columnwidth}\textbf{d)}\\
\includegraphics[width=0.49\columnwidth]{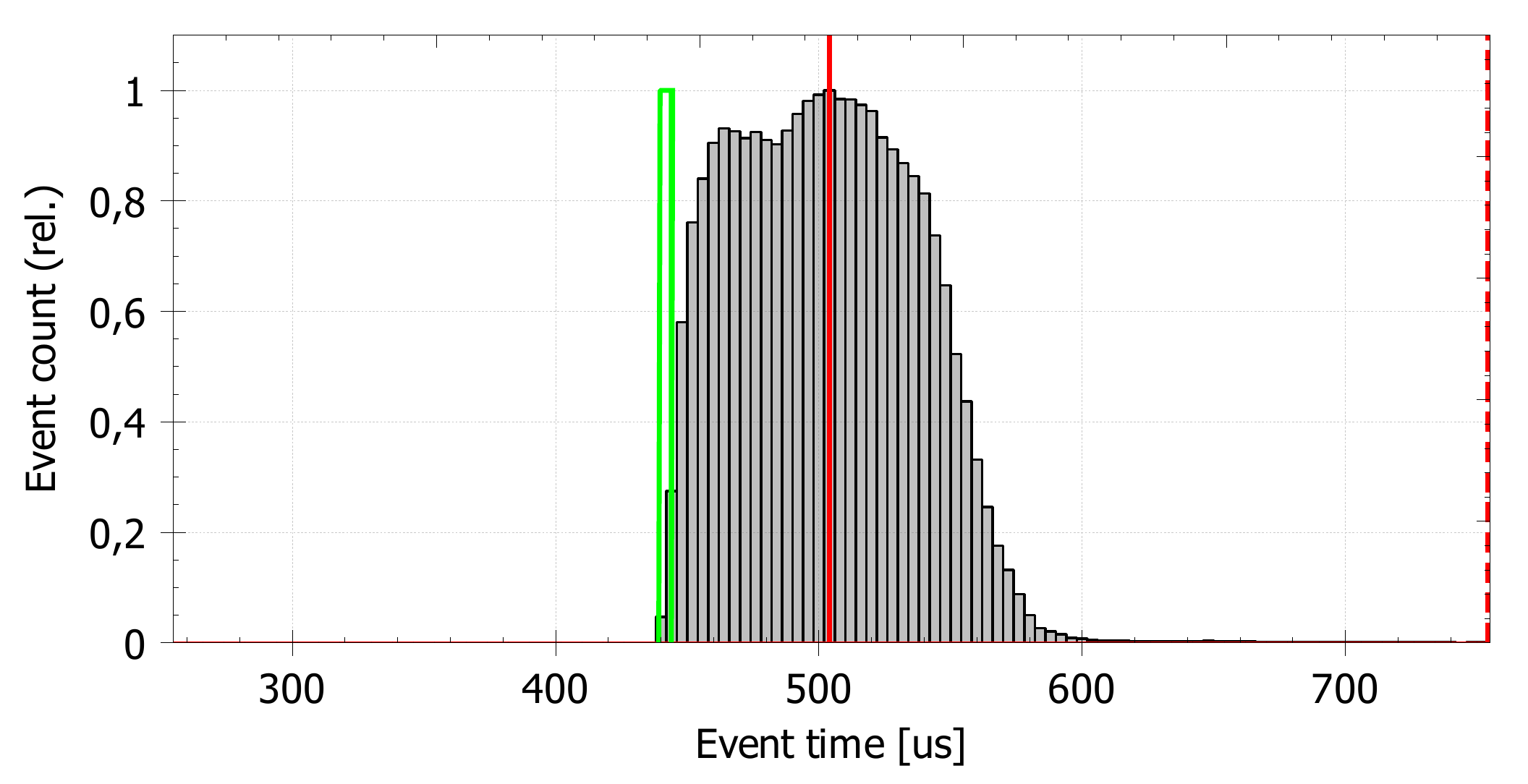}
\includegraphics[width=0.49\columnwidth]{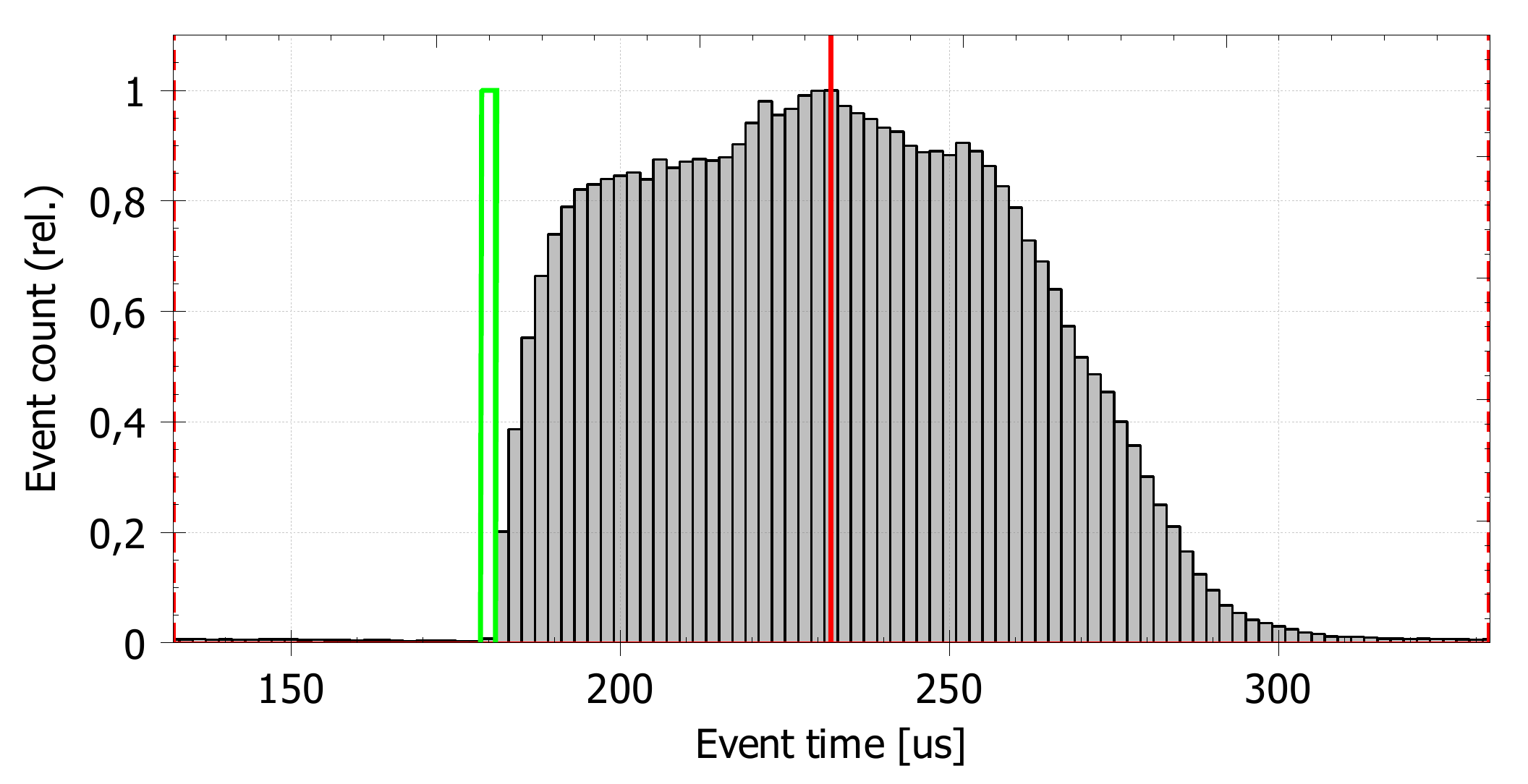}
\caption{Histograms of recorded event times for pulsed-EBIV measurements on a water jet at pulsing frequencies of $f_p \!=\! 2$\,kHz (\textbf{a,c}) and $f_p \!=\! 5$\,kHz (\textbf{b,d}). The green lines indicate the actual time and width of the pulses driving the laser. Top row (\textbf{a,b}) shows a random sample within the record whereas the lower row (\textbf{c,d}) provides the cumulative distribution of events for the respective pulsing frequencies.
The vertical red line marks the peak in the event distribution.} \label{fig:event-hist-waterjet}
\end{figure}

Histograms of the recorded event data are presented in Fig~\ref{fig:event-hist-waterjet} for two laser pulsing frequencies, $f_p \!=\! 2$\,kHz and $f_p \!=\! 5$\,kHz.
The event histograms in Fig.~\ref{fig:event-hist-waterjet}a,b exhibit distinct peaks at the laser modulation frequency with the pulses indicated by the green bursts.
Cumulative histograms compiled from a larger number of pulse periods (Fig.~\ref{fig:event-hist-waterjet}c,d) provide a temporal distribution of the event generation due to pulsed light.
A strong increase in events coincides with the beginning of the laser pulse and peaks after about $100\,\mu$s before decaying.
Most events associated with the laser pulse are captured within $150-200\,\upmu$s.
This time scale determines the maximum laser pulsing rate at which reliable pulsed-EBIV can be obtained.
For the present event camera (Prophesee EVK2-HD) this corresponds to about 5--7\,kHz when recording at full sensor resolution ($1280 \!\times\! 720$) and the arbiters operate below their saturation (i.e. less than $40 \!\cdot\! 10^6$ events/s).

\begin{figure}[htb]
\small
\textbf{a)}\hspace{0.49\columnwidth}\textbf{b)}\\
\includegraphics[width=0.49\columnwidth]{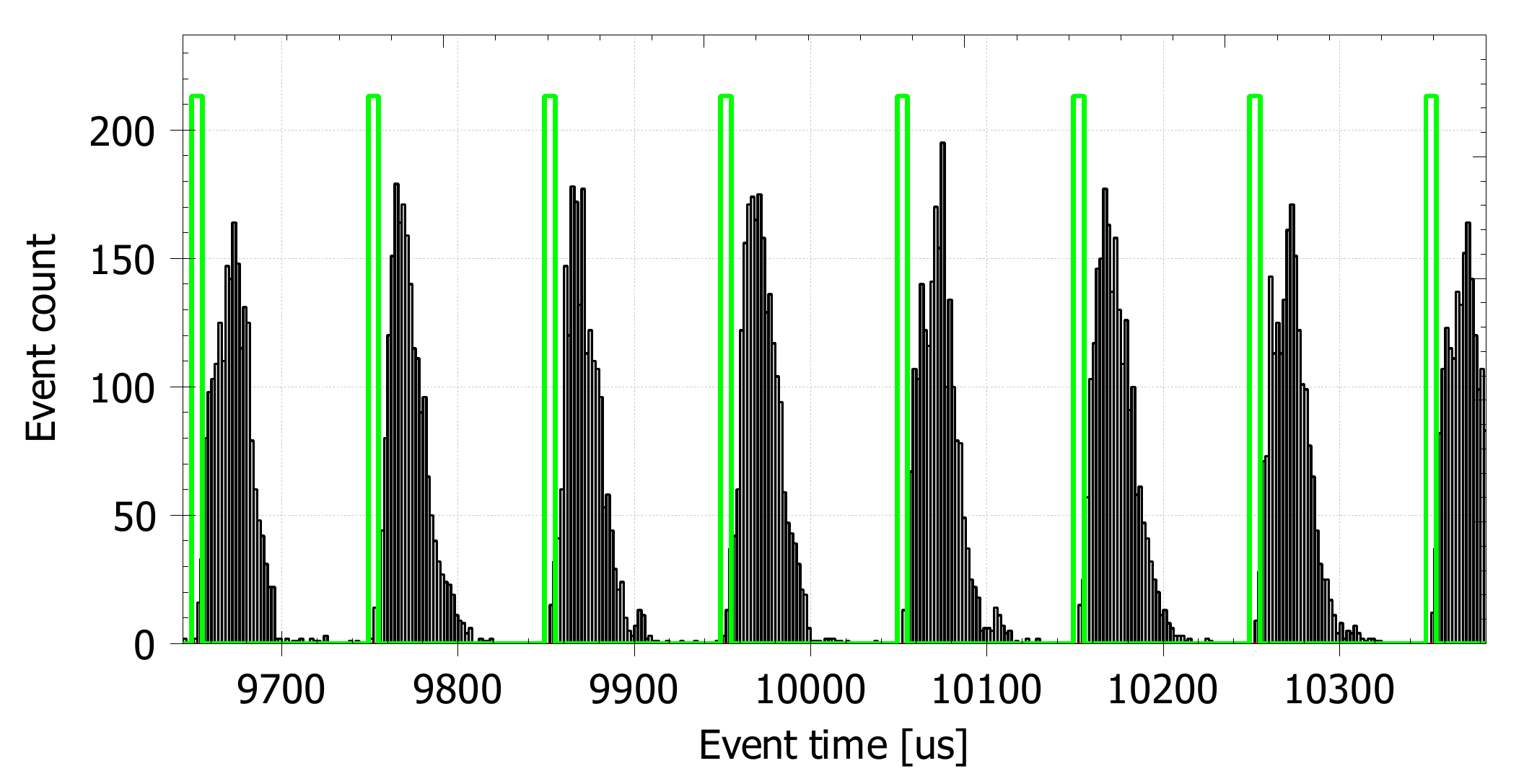}
\includegraphics[width=0.49\columnwidth]{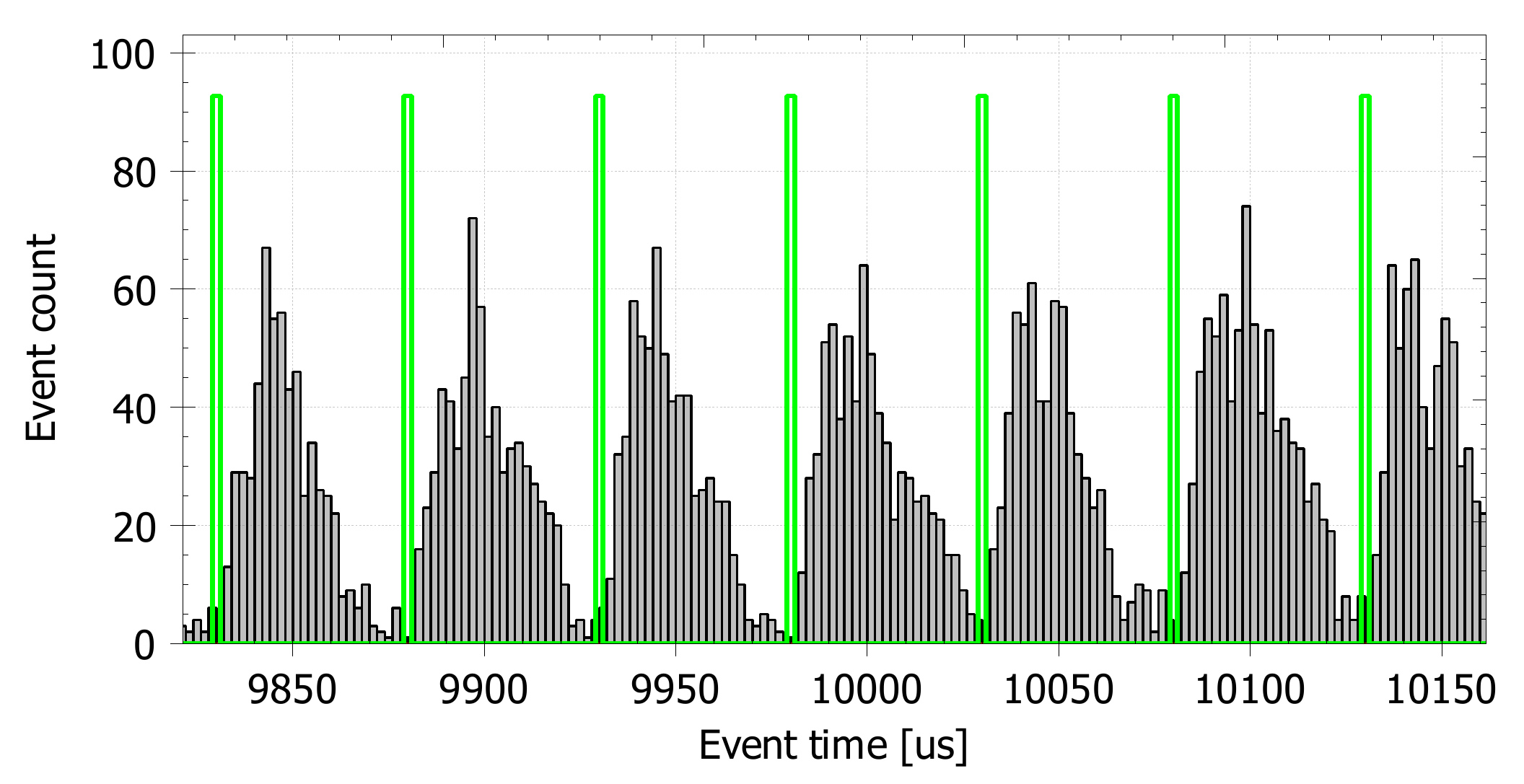}\\
\textbf{c)}\hspace{0.49\columnwidth}\textbf{d)}\\
\includegraphics[width=0.49\columnwidth]{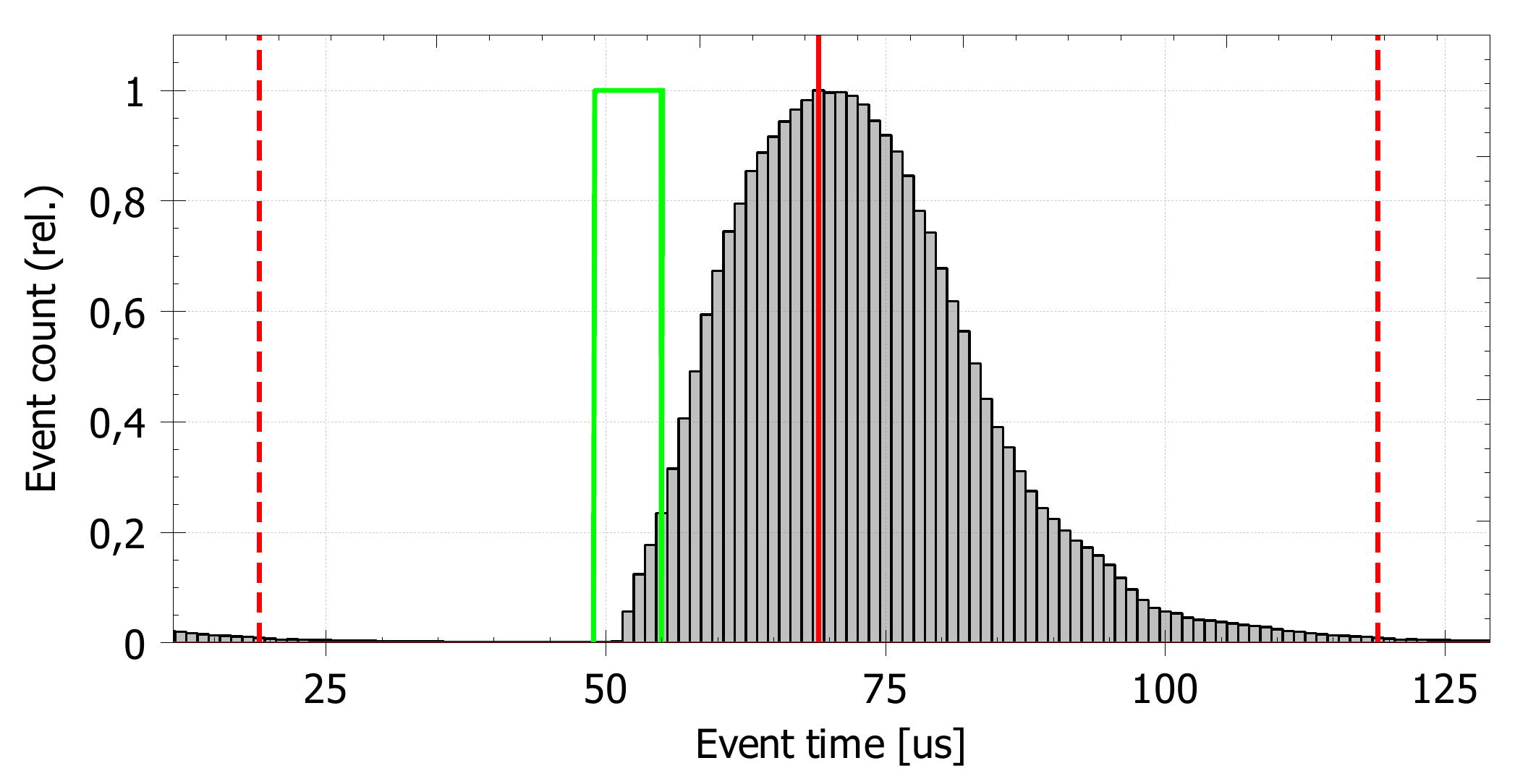}
\includegraphics[width=0.49\columnwidth]{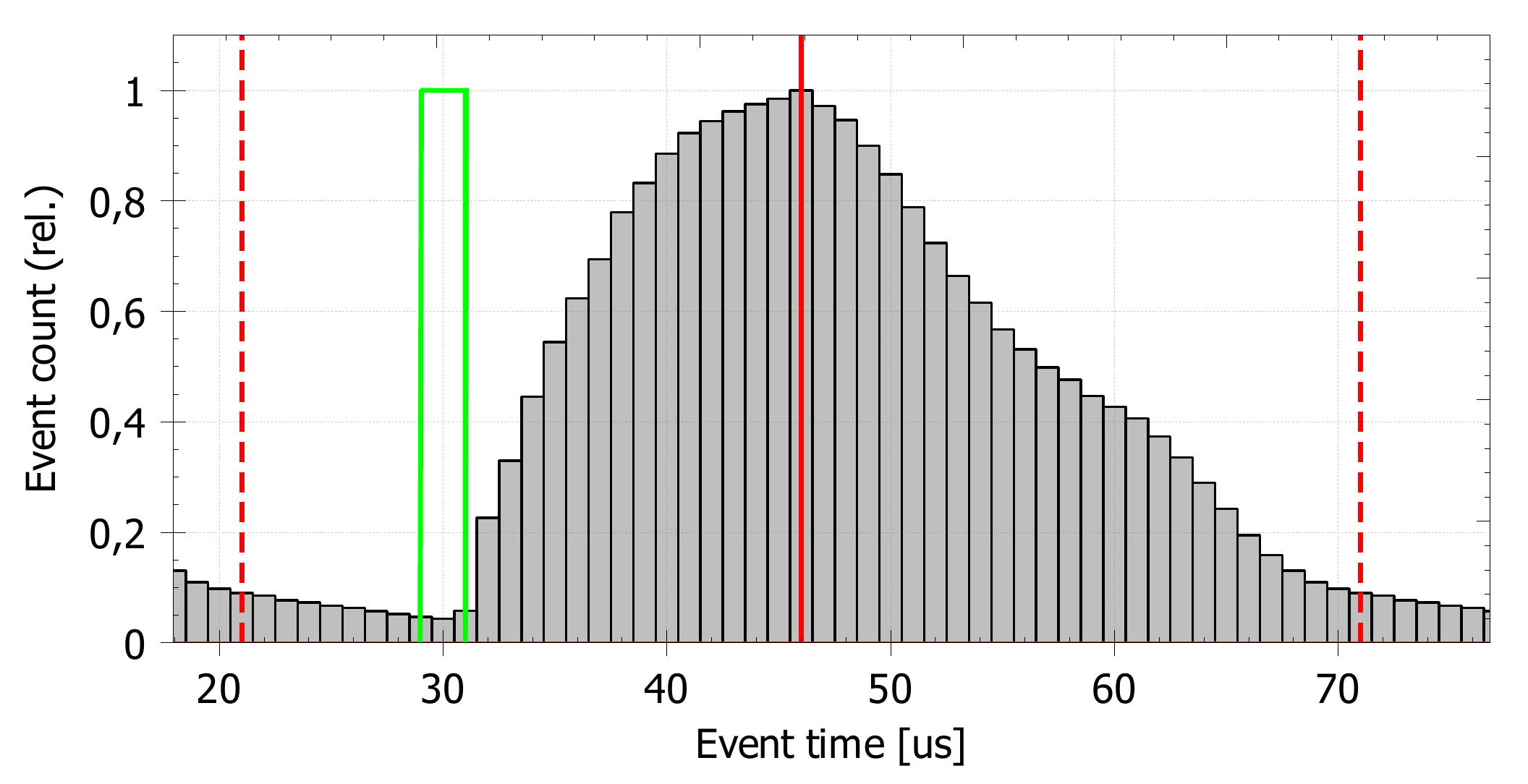}
\caption{Histograms of recorded event times in a ROI of $1280 \!\times\! 160$ pixel for laser pulse frequencies of $f_p \!=\! 10$\,kHz (\textbf{a,c}) and $f_p \!=\! 20$\,kHz (\textbf{b,d}).  Pulse widths are 6\,$\upmu$s and 3\,$\upmu$s respectively producing event pulses widths of 26\,$\upmu$s and 22\,$\upmu$s (FWHM).
Top row shows sequences of several pulses, bottom row a cumulative distribution from several hundred pulse periods.  The dashed red vertical lines mark the pulsing period duration.
Green line indicates pulse driving the laser, actual pulse of light may be longer (c.f. Fig.~\protect\ref{fig:laser-pulses}).
}
    \label{fig:event-hist-roi}
\end{figure}

By reducing the active area of the sensor (\ac{ROI}), the performance of the arbiter can be enhanced, as the number of events per row and unit time is reduced while keeping the particle image density within the ROI constant.
Fig.~\ref{fig:event-hist-roi} shows histograms of event data obtained for a \ac{ROI} of 1280($W$) $\times$ 160($H$) pixel with corresponding event pulse widths of 26\,$\upmu$s and 22\,$\upmu$s (\ac{FWHM}).
This indicates that the motion of particles can be captured at pulsing frequencies up to 20\,kHz at this \ac{ROI}.

\begin{figure}[htb]
\small
\textbf{a)}\hspace{0.45\columnwidth}\textbf{b)}\\
\includegraphics[width=0.49\columnwidth]{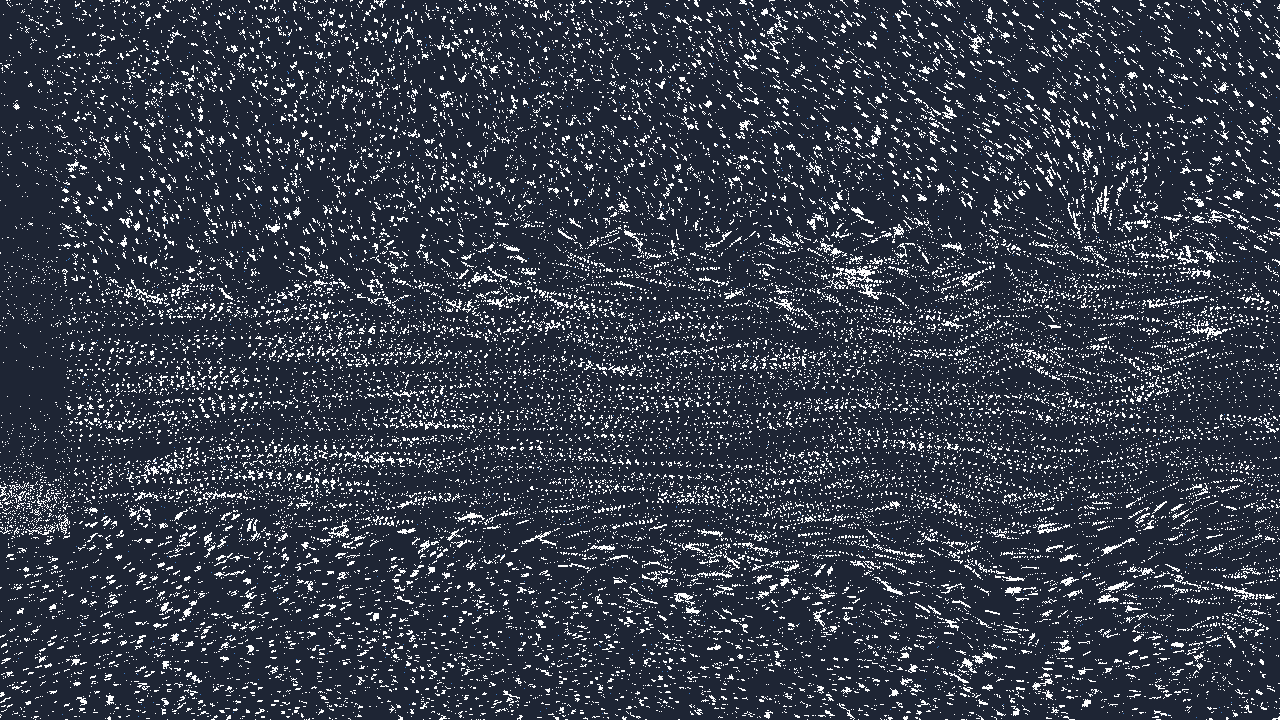}
\includegraphics[width=0.49\columnwidth]{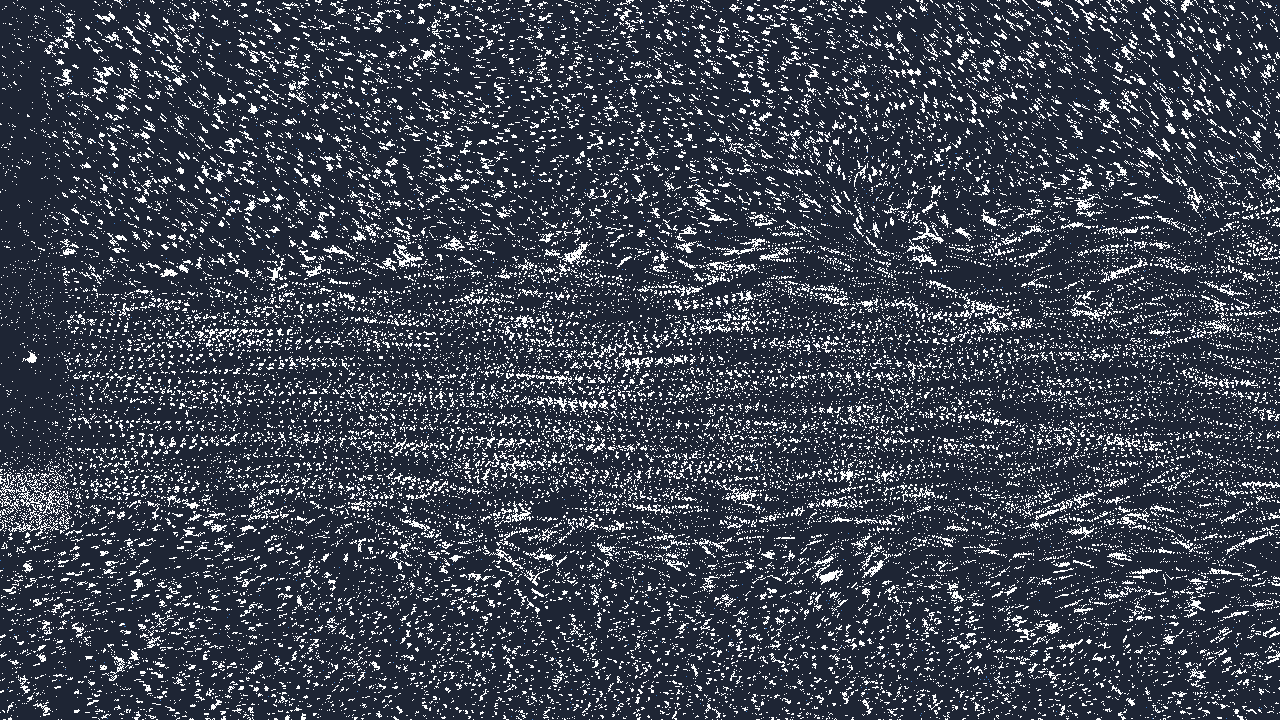}\\
\textbf{c)}\hspace{0.45\columnwidth}\textbf{d)}\\
\includegraphics[width=0.49\columnwidth]{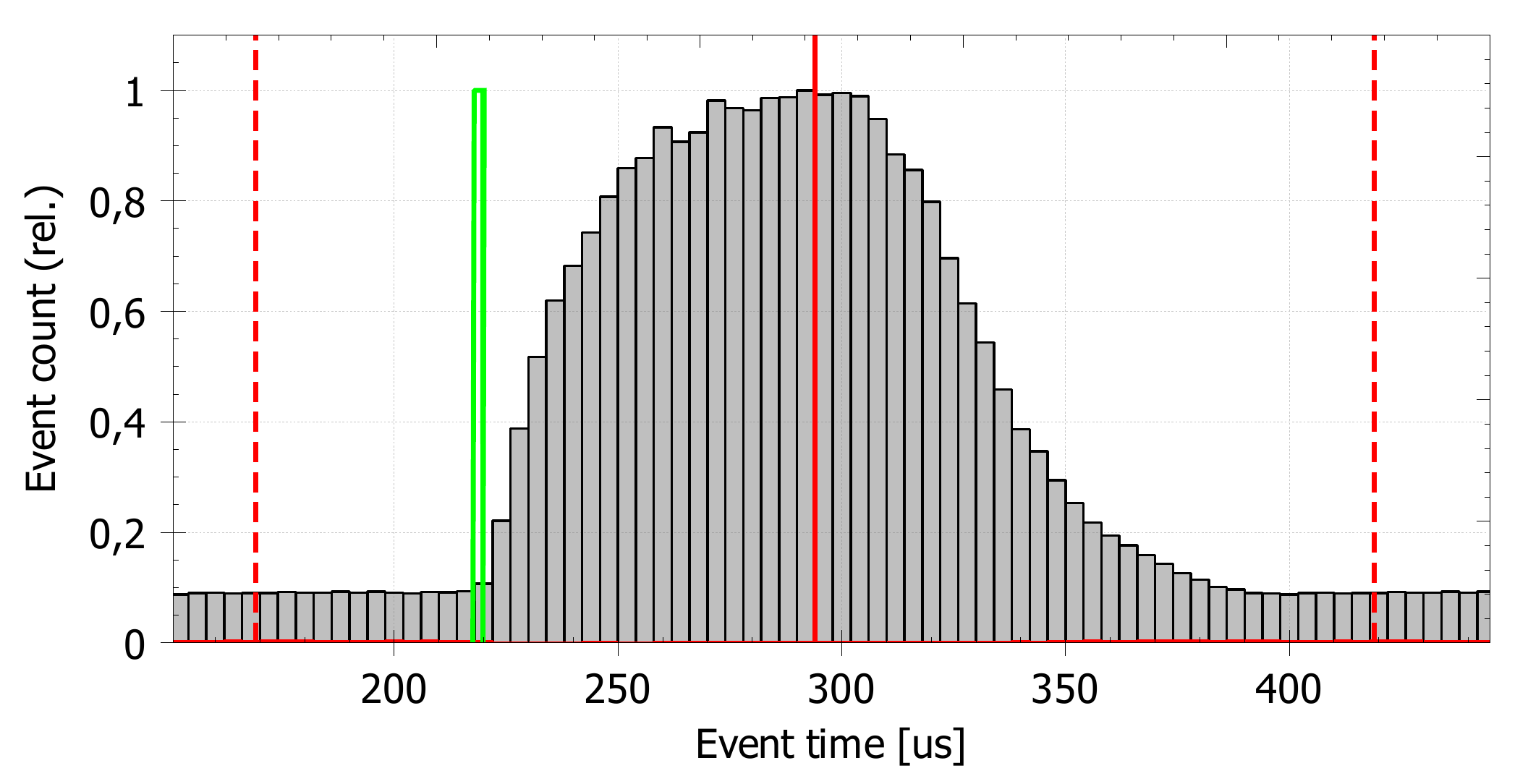}
\includegraphics[width=0.49\columnwidth]{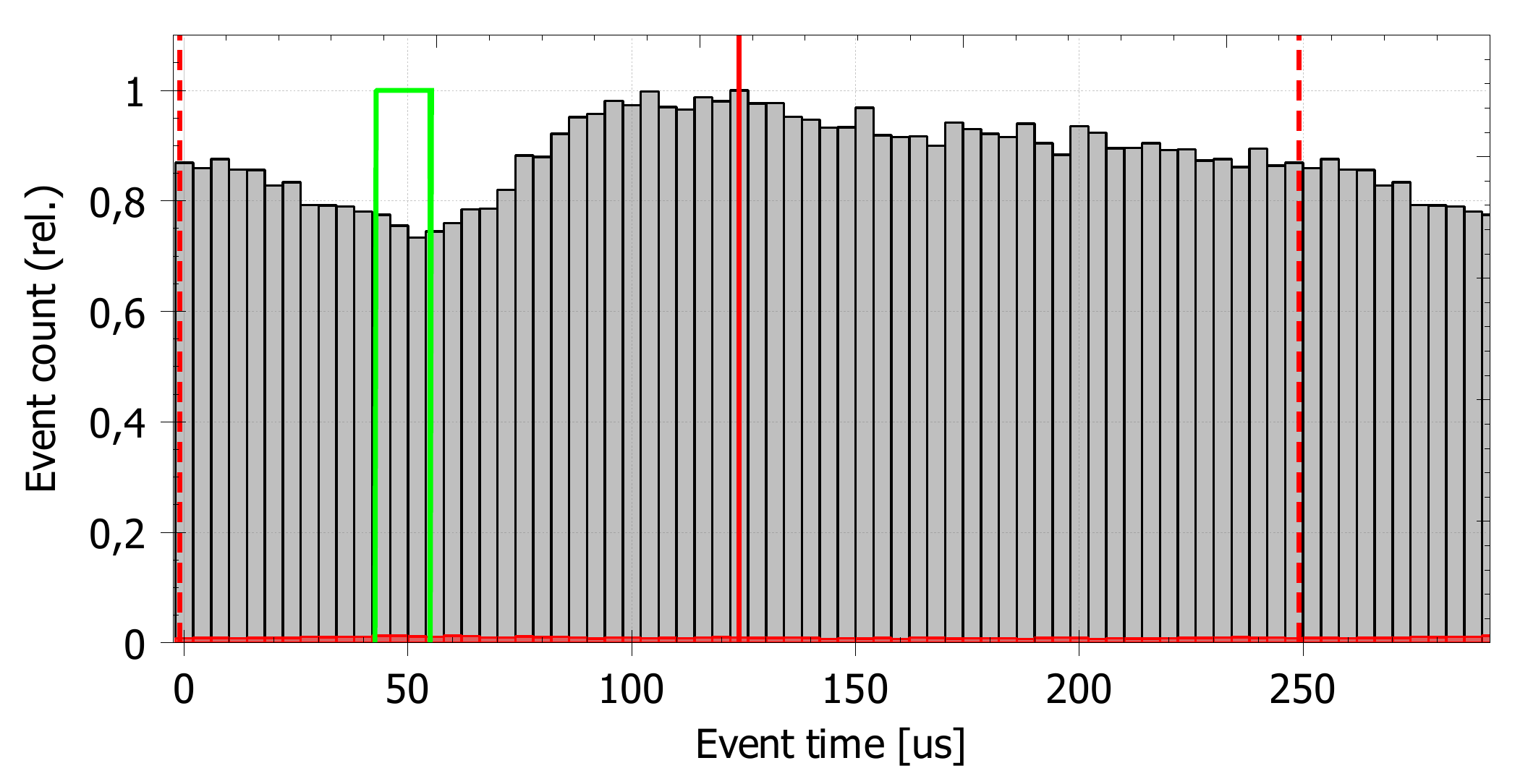}
\caption{Sample of 2.5\,ms of events recorded on a turbulent jet in water at $f_p \!=\! 4$\,kHz and event rates of $38 \!\cdot\! 10^6$ events/s (\textbf{a}) and $50 \!\cdot\! 10^6$ events/s (\textbf{b}). Subfigures (\textbf{c}) and (\textbf{d}) show respective temporal distributions of the events. The dashed red vertical lines indicate one laser pulsing period ($250\,\upmu$s).
	(Animations of the event data in (\textbf{a}) are provided in the supplementary material.)
} \label{fig:waterjet-saturated}
\end{figure}

The effect of increasing the event data rates is highlighted in Fig.~\ref{fig:waterjet-saturated}.
While the event pseudo-images in Fig.~\ref{fig:waterjet-saturated}a,b look nearly the same, the corresponding event histograms (Fig.~\ref{fig:waterjet-saturated}c,d) reveal that the clear defined burst of events following a laser pulse disappears when the event rate increases from $38 \!\cdot\! 10^6$ events/s to $50 \!\cdot\! 10^6$ events/s.
Due to arbiter saturation, the event data in Fig.~\ref{fig:waterjet-saturated}b cannot be reliably processed to extract particle image displacement data.
Online monitoring is therefore required during event capture to ensure adequate recordings.

\begin{figure}[htb]
\small
%\textbf{a)}\hspace{0.49\columnwidth}\textbf{b)}\\
\includegraphics[trim=0 0 60 0,clip,width=\columnwidth]{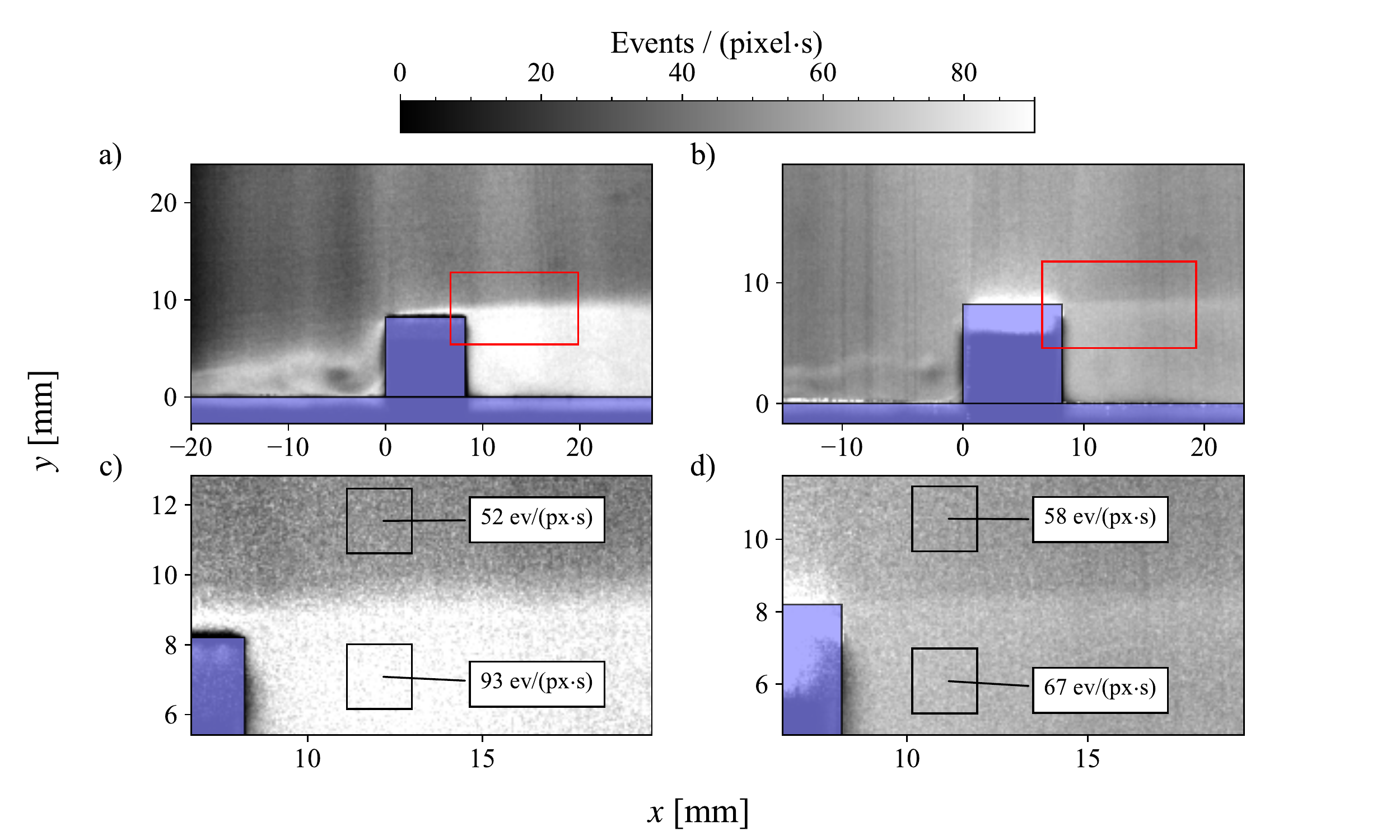}
\caption{Event rate for the flow around a square rib for continuous illumination (\textbf{a,c}) and pulsed illumination at 5~kHz (\textbf{b,d}). Red rectangles in (\textbf{a,b}) indicate region of interest shown in (\textbf{c,d}).}
    \label{fig:rib-event-rate}
\end{figure}

As mentioned previously, continuous illumination results in event data rates that strongly depend on the particles' velocity.
This is illustrated in Fig.~\ref{fig:rib-event-rate}a,c for an air flow around a square rib (c.f. Sect.~\ref{sec:rib-flow}) with data rates differing by nearly a factor of two between fast outer flow and the slow flow in the wake region.
The seeding distribution on the other hand is constant throughout the imaged domain.
With pulsed illumination, here at 5\,kHz, the event generation is nearly uniform in all areas of the imaged domain.

\section{Processing of pulsed event-data}
\label{sec:ebiv-processing}
Event-data provided by the pulsed illumination of the particles suggests the use of established multi-frame \ac{PIV} processing algorithms.
This is achieved by first generating pseudo-images from the event data. These are accumulated from the events surrounding a given maximum in the event histograms as shown in Fig.~\ref{fig:event-hist-waterjet} and can be readily determined if the pulsing frequency is known. 
An optimal association of events is possible when the timing of the laser pulses has been recorded as part of the event stream.
All events within a given frame are assigned with the same time stamp.

Processing the event sequences provided by the pulsed illumination involves several steps in order to convert the event data stream into distinct pseudo images that can be handled by typical correlation-based algorithms designed for handling \ac{PIV} recordings:
\begin{enumerate}
    \item Using the known, fixed pulsing frequency estimate the temporal position of the beginning of the light pulse (in case the laser trigger pulses are not recorded by the event camera).
    \item Generate sequence of pseudo images using all events for a given light pulse. Pixels associated with a \ac{CD} event are given the same intensity irrespective of their actual time-stamp under the assumption that the given pixels were triggered by the same light pulse. The resulting stream of images will have the same frequency as the laser pulses used to illuminate the particles.
    \item Image blurring is suggested to reduce spiking of the cross-correlation signals and ensure a correlation peak width suitable for sub-pixel peak position estimation. A smoothing filter with a kernel size of $1-2$\,pixels is considered adequate.
\end{enumerate}
As such modern \ac{PIV} algorithms are optimized for particle image intensity distributions of approximately Gaussian shape and perform less well with the single-pixel events (Dirac delta function).
Therefore, the pseudo-images generated from the event data are first lowpass filtered with a Gaussian kernel to generate suitable ``particle images".

The sequence of pseudo-particle image distributions can be processed using standard correlation-based \ac{PIV} algorithms as mentioned above and significant improvement can be achieved through the use of multi-frame, pyramid-based cross-correlation schemes \citep{Sciacchitano:2012,LynchScarano:2013,Willert:ISPIV:2021}.
The analysis of the pseudo-image sequences obtained in the following experiments typically involves sub-sets of $N_f \!=\! 5$ frames per time-step and is increased up to $N_f \!=\! 9$ to reduce noise at the cost of reduced frequency bandwidth due to temporal filtering.

\section{Sample measurements performed with pulsed-illumination EBIV}

\subsection{Turbulent jet in water}
\label{sec:ebiv-waterjet}
Both pulsed-EBIV and matching snapshot PIV measurements are performed on a turbulent jet in a small water tank as described in Sect.~\ref{sec:meas-latency}.
The field of view is adjusted to achieve similar magnification factors for both measurement techniques, $m \!=\! 27.0$\,pixel/mm (37.0\,$\upmu$m/pixel) for pulsed-EBIV and $m \!=\! 28.75$\,pixel/mm (34.8\,$\upmu$m/pixel) for PIV, using the same lens for both (Nikon Micro-Nikkor 55\,mm 1:2.8).
Laser light sheet configuration, seeding concentration and operational conditions of the jet are held constant to achieve a common imaging configuration.
A side-by-side comparison of the two measurements is given in Table~\ref{tbl:piv_ebiv}.

For PIV, the laser is operated in double-pulse mode with a pulse separation of $\tau \!=\! 500\,\upmu$s and pulse duration of $t_p \!=\! 100\,\upmu$s ($600\,\upmu$J). PIV recordings are captured at about 4~Hz using a double-shutter CCD camera (PixelFly-PIV, PCO/ILA-5150) with a resolution of $1392 \!\times\! 1040$ pixels at $6.45\,\upmu$m/pixel.
The acquired \ac{PIV} data set comprises 1000 image pairs which are processed with a standard multiple-pass, iterative grid-refining cross-correlation based algorithm.
As an initial step the mean intensity image is computed from the image set and subtracted from each image prior to processing, with negative intensity values set to zero.
The final sampling window size covers $32 \!\times\! 32$\,pixels ($1.11 \!\times\! 1.11\,\mathrm{mm}^2$). Validation is based on normalized median filtering with a threshold of 3 \citep{WesterweelScarano:2005}.

For pulsed-EBIV, the laser is operated with pulsing frequencies of 2,4 and 5 kHz and a duty cycle of 1\% while the camera lens aperture is stepped down to $f_\# \!=\! 5.6$. Pulse energies vary in the range of about $10-25\,\upmu$J, depending on pulsing rate. This is a small fraction of the energy used for PIV. The event camera biases are adjusted for maximum sensitivity to positive \ac{CD} events given that negative \ac{CD} events are not required in the subsequent processing. These parameters result in event data rates of $35-38 \!\cdot\! 10^6$ events/s (100--120 MB/s).
Additional data is acquired from a reduced \ac{ROI} of $320 \!\times\! 720$\,pixel at similar effective data rates and pulsing frequencies up to 10 kHz.
Event sequences of 10\,s duration are recorded for both sensor \ac{ROI}s.

Processing of the event recordings first involves the conversion to pseudo-image sequences as described in Sect.~\ref{sec:ebiv-processing} followed by displacement field estimation using a multi-frame cross-correlation scheme.
The iterative $N_f \!=\! 5$-frame algorithm uses sampling windows of $48 \!\times\! 48$\,pixel for event data acquired on the full HD array ($1280 \!\times\! 720$\,pixel) and is reduced to $32 \!\times\! 32$\,pixel for a \ac{ROI} of $320 \!\times\! 720$\,pixel.
Additionally, the full HD event data obtained at 5\,kHz is processed at $32 \!\times\! 32$\,pixel with a $N_f \!=\! 9$-frame scheme.

\begin{figure}[htb]
\small
\textbf{a)}\hspace{0.45\columnwidth}\textbf{b)}\\
\includegraphics[width=0.49\columnwidth]{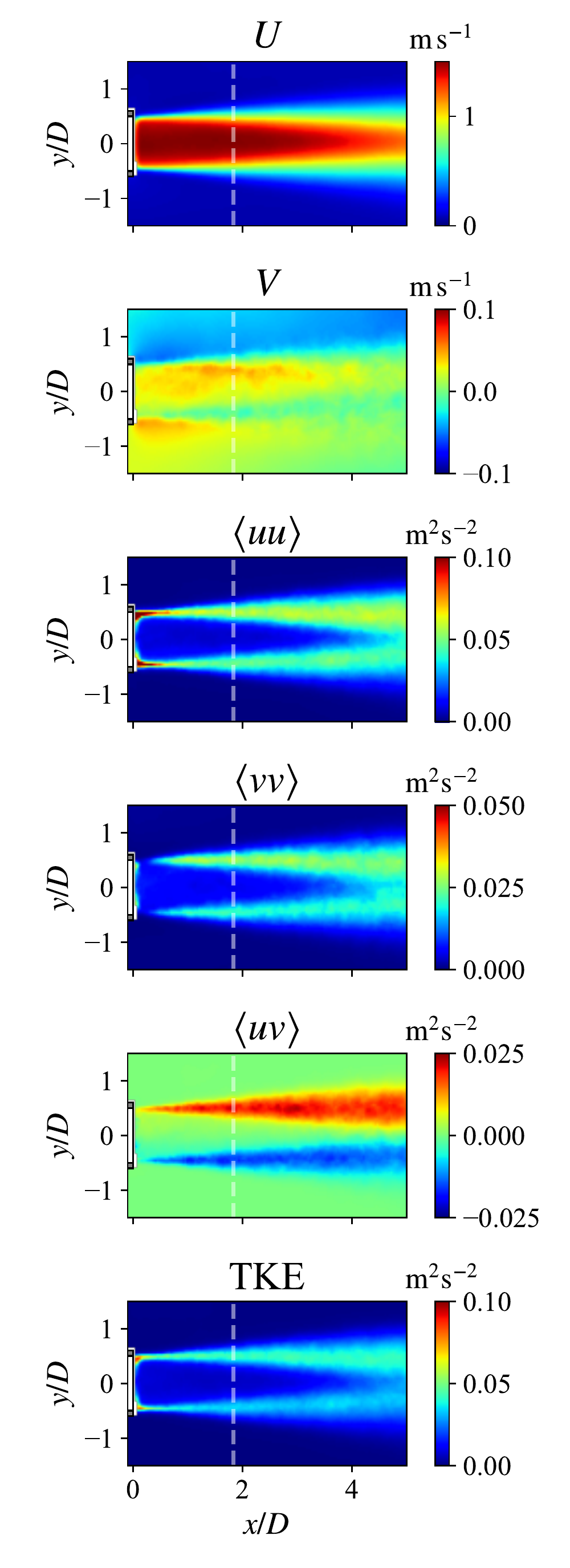}
\includegraphics[width=0.49\columnwidth]{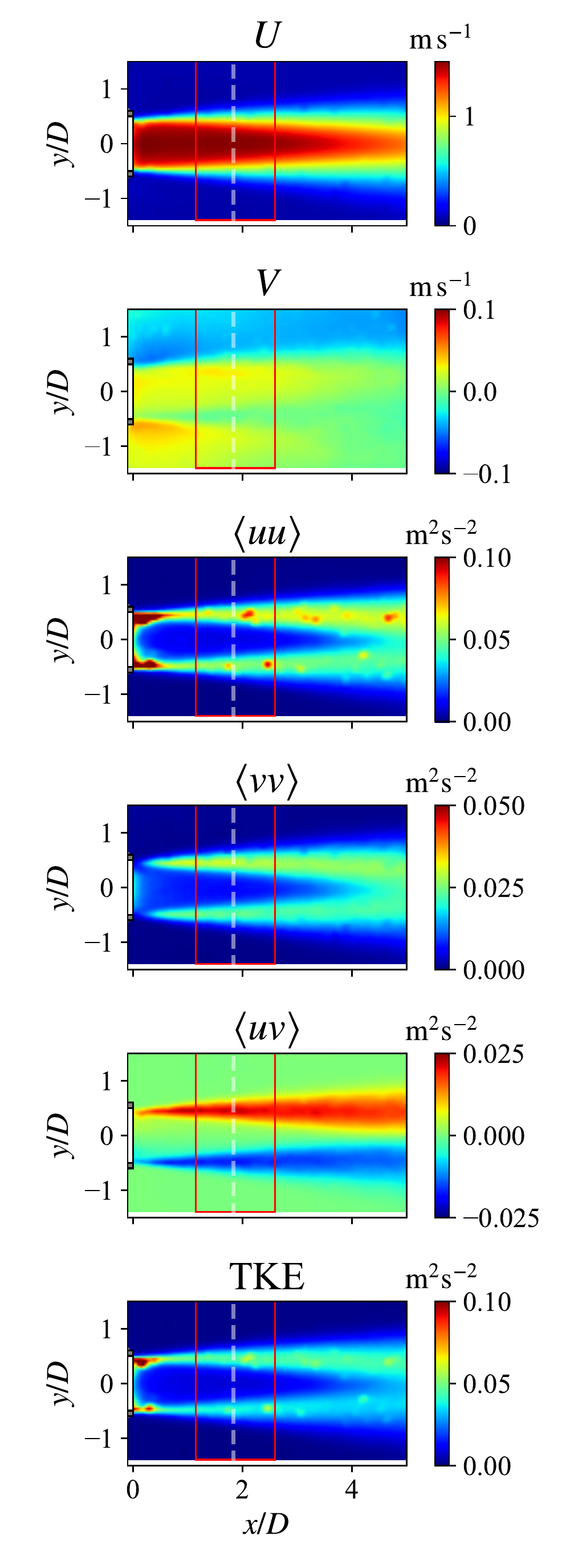}
\caption{Maps of velocity statistics obtained with PIV (a) and pulsed-EBIV (b) with a pulsing frequency of 5\,kHz.
Red box shows ROI used for increased event rate capture;
dashed line at $x/D \!=\! 2$ indicates profile sampling position.} \label{fig:waterjet-stats}
\end{figure}

\begin{figure}[htb]
\small
\textbf{a)}\hspace{0.45\columnwidth}\textbf{b)}\\
\includegraphics[width=0.49\columnwidth]{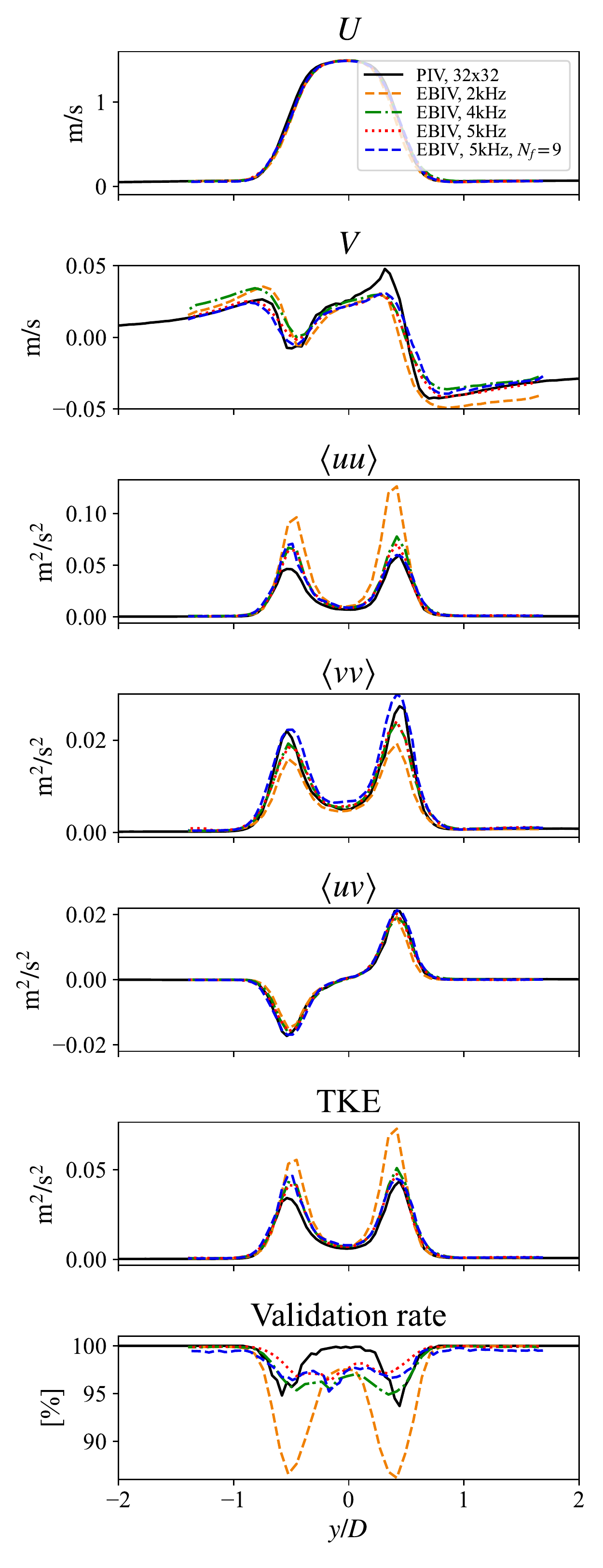}
\includegraphics[width=0.49\columnwidth]{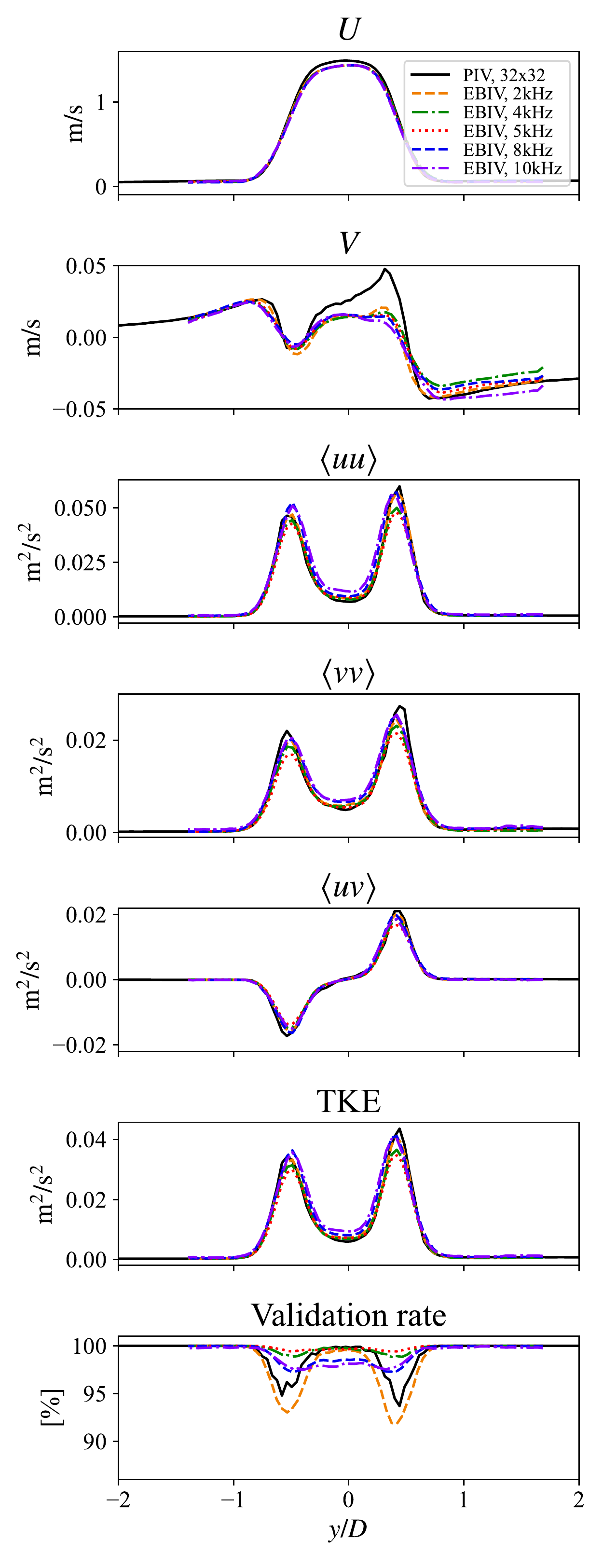}
\caption{Profiles of velocity statistics at $x/D \!=\! 2$ obtained with PIV  and pulsed-EBIV at different pulsing frequencies, for full-field event imaging at $1280 \!\times\! 720$ pixels (a) and limited region of interest at $320 \!\times\! 720$ pixels (b) around $1 < \! x/D\! < 2.8$.} \label{fig:waterjet-profiles}
\end{figure}

Fig.~\ref{fig:waterjet-stats} provides a side-by-side comparison of the first and second order statistics for both PIV and EBIV data.
Except for the region near the nozzle there is good quantitative agreement between the two data sets.
Asymmetry in the transverse velocity $v$ is due to wall effects within the small water tank. 
Also a deflecting plate located further downstream (at $x/D \!\approx\! 20$) was moved between the experiments thereby slightly altering the cross-flow.
The streamwise velocity variance $\langle u u \rangle$ shows some irregular structures the cause of which have not been identified.
Profiles extracted at $x/D \!=\! 2$ are shown in Fig.~\ref{fig:waterjet-profiles} and exhibit a good collapse for the mean value. 
The largest discrepancies are present for the variances $\langle u u \rangle$ and $\langle v v \rangle$ for the event data acquired at $f_p \!=\! 2$\,kHz.
An explanation for this is that the $N_f \!=\! 5$-frame processing scheme insufficiently captures the flow dynamics in the jet's shear layers.
This is also reflected in the validation rate which drops considerably in the shear layers for the $f_p \!=\! 2$\,kHz event data.

Considerable improvement in the convergence of the different data sets is achieved through the reduction of the \ac{ROI} (Fig.~\ref{fig:waterjet-profiles}b). 
This is attributed by the increase of the particle image density by roughly a factor of 3--4 (c.f. Table~\ref{tbl:ebiv_measdata}. At the same time the spatial resolution could be increased by decreasing the sample size from $48 \!\times\! 48$ pixel to $32 \!\times\! 32$ pixel.

\begin{table}
  \caption{Comparison of snapshot PIV and pulsed EBIV at 4~kHz on a jet flow in water. Light sheet and seeding density is held constant.} \label{tbl:piv_ebiv}
  \centering
\begin{tabular}{lcc}
  \hline
   & \textbf{PIV} & \textbf{pulsed EBIV} \\[2pt]
  \hline
  Illumination & double pulses @ 4~Hz & 4~kHz \\
  Pulse delay, $\tau_p$ & $500\,\upmu$s & $250\,\upmu$s \\
  Pulse width & $100\,\upmu$s & $7.5\,\upmu$s \\
  Lens aperture & $f_\# 2.8$ & $f_\# 4.0$ \\
  Magnification & $34.8\,\upmu$m/pixel & $37.0\,\upmu$m/pixel \\
  Recording duration & $\approx 4$\,min & 10\,s \\
  Data set size & 1000 image pairs (12\,bit) & 40\,000 pseudo images (1\,bit) \\
     & 3.25\,GB & 0.95\,GB \\
  Frames used per time-step & $N_f = 2$ & $N_f = 5$ \\
  Interrogation sample size & $32\times 32$ & $48\times 48$ \\
         & $1.1\times 1.1\,\mathrm{mm}^2$ & $1.8\times 1.8\,\mathrm{mm}^2$\\
  Particle image density$^1$, $ppp$ & 0.0088 & 0.0033 \\[2pt]
\hline
\end{tabular}
\small{$^1$) Details on $ppp$ estimation are provided in appendix section~\ref{sec:Appx:seeding density}.}
\end{table}

%--------------------------------------------------------------
\subsection{Channel boundary layer}
\label{sec:tbl}
Measurements of a developing boundary layer are performed in the square duct of a small wind tunnel at $x/D \!\approx\! 20$ with $D \!=\! 76$\,mm being the cross section dimension.
Depending on the flow velocity the flow switches from laminar to turbulent conditions with the transition occurring at about $U_\infty \!=\! 3.0$\,m/s.
It should be noted that at $x/D \!\approx\! 20$ (1475 mm) the turbulent duct flow is not fully converged.

Reduced sensor \ac{ROI}s of $320(W) \!\times\! 720(H)$ and $1280(W) \!\times\! 320(H)$ capture a $7$\,mm wide strip of the flow field respectively extending up to wall normal distances of $y \!= \!15$\,mm ($y/D = 0.39$) and $y \!=\! 27$\,mm ($y/D \!=\! 0.73$).
Respective magnification is  $m \!=\! 46.6$\,pixel/mm for the laminar condition and $m \!=\! 45.9$\,pixel/mm for the turbulent condition.

Under laminar conditions the imaged particles produce event clusters with a mean size of 1.29 events per particle and light pulse. The particle image density, expressed in \ac{ppp}, is estimated at $ppp \!=\! 0.014$ and includes a small amount (O[1\%]) of random events, which are triggered randomly on the sensor array (shot noise).
Event data captured of the turbulent channel flow condition exhibit a similar event cluster size (1.45) at a slightly lower image density of $ppp \!=\! 0.011$. (Details on the estimation of particle image density and size are provided in  appendix~\ref{sec:Appx:seeding density}.)

\begin{SCfigure}[\sidecaptionrelwidth][htb]
\small
\includegraphics[width=0.6\columnwidth]{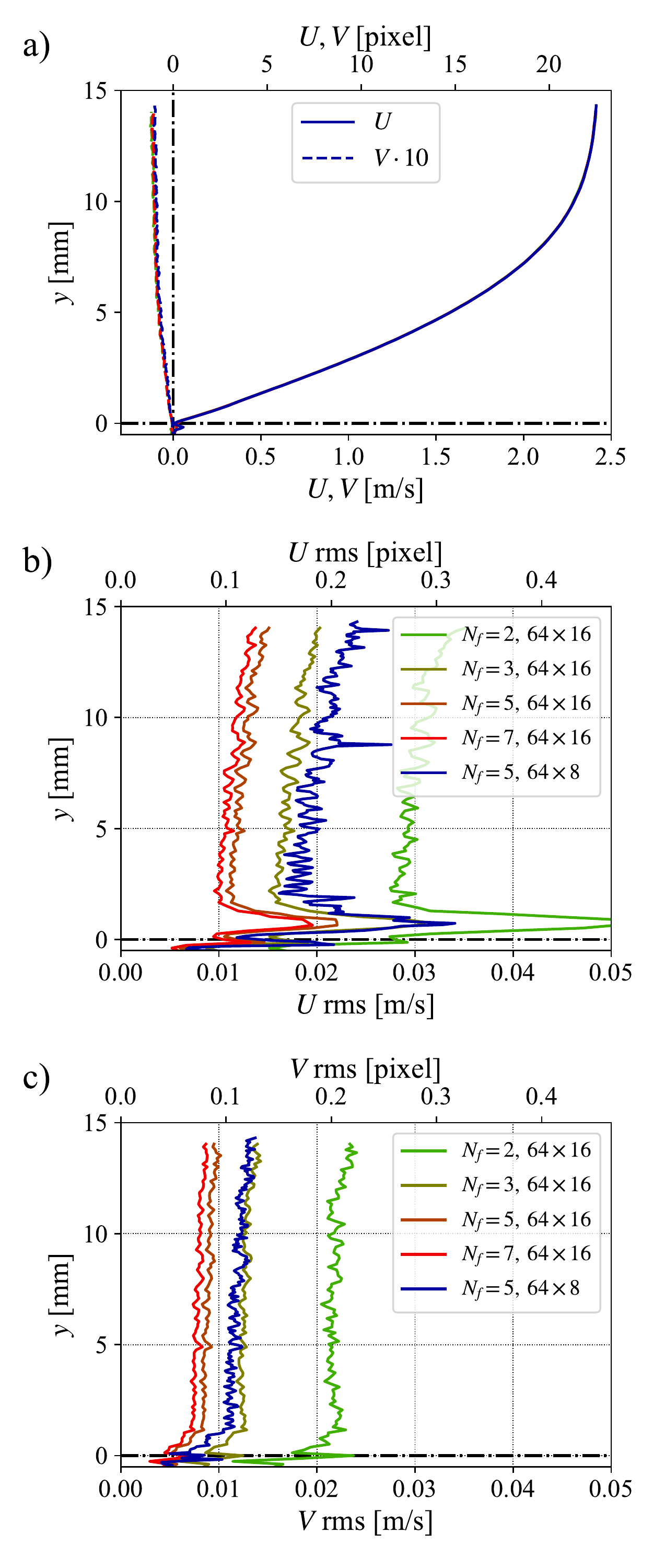}
\caption{Mean velocity profile for a laminar channel boundary layer obtained with pulsed light EBIV (\textbf{a}); profiles of the rms values of streamwise (\textbf{b}) and wall-normal component (\textbf{c}). Particle image density is estimated at $ppp\!=\! 0.014$ with 15 particles per $64\!\times\!16$ pixel sample.}
    \label{fig:lambl_profile}
\end{SCfigure}

The velocity profile of the laminar flow condition determined from an event stream at a laser pulse rate of $f_p = 5\,000$\,Hz and 1.0\,s duration is shown in Fig.~\ref{fig:lambl_profile}a.
The \ac{rms} values of the velocity components (Fig.~\ref{fig:lambl_profile}b,c) are determined by subtracting a moving average of $N \!=\! 50$\,samples (10\,ms) from the time-resolved velocity profile to account for a slight unsteadiness of the tunnel mean flow.
With the low frequency part removed, the \ac{rms} values give an indication of the measurement uncertainty that can be achieved with the present pulsed-EBIV implementation.
While the choice of interrogation window size and number of frames, $N_f$, used per time-step have insignificant influence on the mean values, they do affect the \ac{rms} values.
Increasing the sample size reduces the \ac{rms} values as more particle image event matches enter into the correlation estimate.
The same can be achieved by increasing $N_f$.
In that case, the underlying sum of correlation scheme attenuates noise in the correlation plane from which the particle image displacement is determined.
At the same time the \ac{rms} values are only weakly correlated with increasing displacement values which is to be expected from the iterative processing scheme that offsets the sample windows according to the displacement \citep{Westerweel:1997,Westerweel:2002}.
The peak in the \ac{rms} value of the streamwise velocity component $u$ close to the wall near $y \!=\! 1$\,mm is of physical nature and is associated with the onset of weak instabilities of the near wall flow.

\begin{figure}[htb]
	\small
	\centering
	\includegraphics[width=\columnwidth]{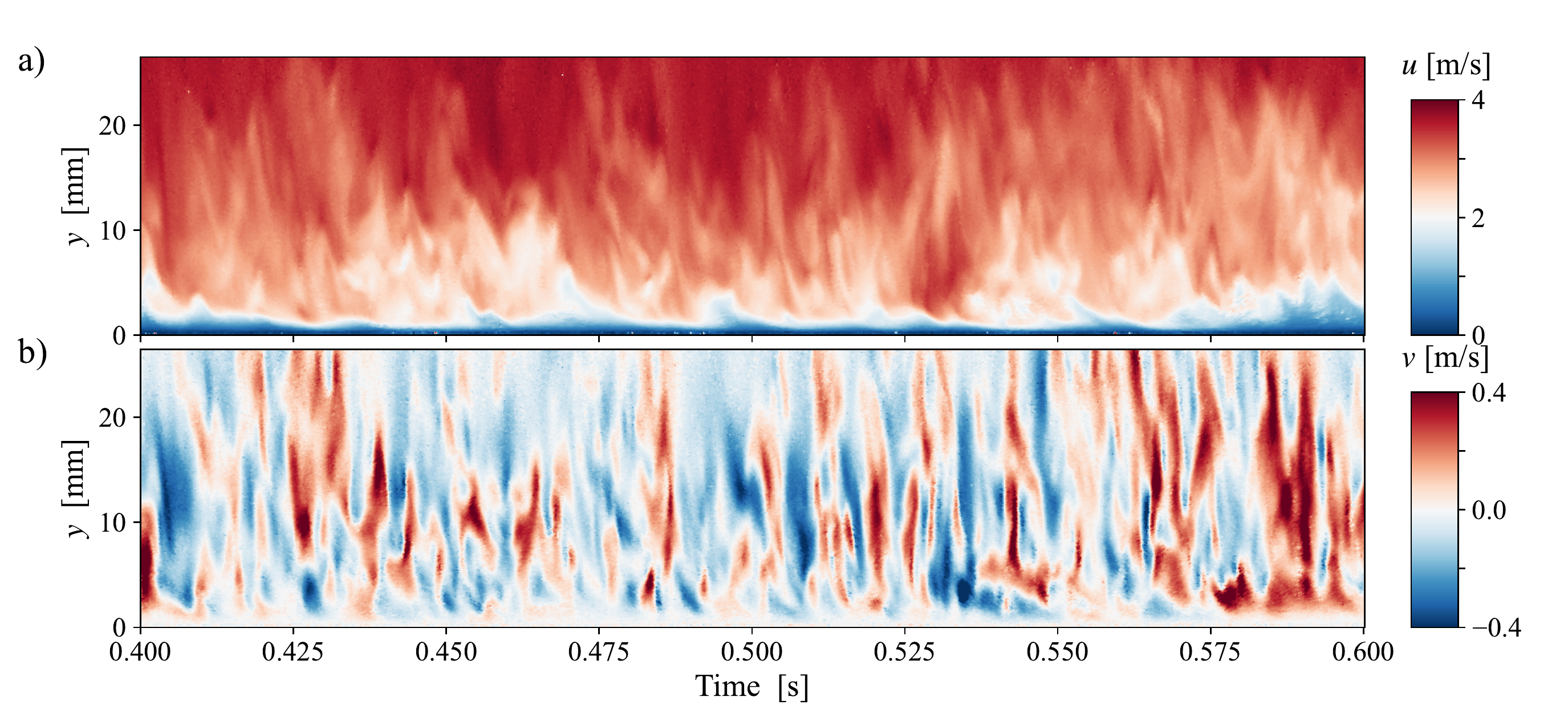}
	\caption{Time-evolving velocity profile covering 0.2\,s (1000 samples) of the streamwise velocity component $u(y,t)$ (\textbf{a}) and wall-normal velocity component $v(y,t)$ (\textbf{b}) for a turbulent channel boundary layer obtained with pulsed light EBIV at 5\,kHz.}\label{fig:turbbl_sequence}
\end{figure}

At slightly higher free stream velocities ($U_\infty\!>\! 3$\,m/s) the channel flow transitions to turbulence. 
Event recordings are performed using the same imaging conditions as for the laminar case at $f_p \!=\! 5$\,kHz and a \ac{ROI} of $1280 \!\times\! 320$\,pixel at $m \!=\! 45.9$\,pixel/mm.
Two event records of 10\,s duration are processed with $N_f \!=\! 5$ images per time step resulting in 50,000 (correlated) vector fields each.
Analysis is performed with a high aspect ratio sampling window $64(W)\!\times\!8(H)$ pixel ($1.39\!\times\!0.174\,\mathrm{mm}^2$) to improve spatial resolution in the wall-normal direction; with sampling locations spaced at $\Delta y\!=\!3$\,pixel ($65\,\upmu$m).
For post-processing the profile of the unsteady velocity components $u,v$ is extracted from the center of each 2d velocity map.
This results in temporal records of the unsteady velocity profile $u_i(y,t)$ at a fixed streamwise position ($x_0\!=\!1475$\,mm).   
A 200\,ms sample of this record is provided in Fig.~\ref{fig:turbbl_sequence} for the two velocity components $u_i\!=\! {u,v}$.

\begin{SCfigure}[\sidecaptionrelwidth][htb]
\small
%\textbf{a)}\hspace{0.45\columnwidth}\textbf{b)}\\
\includegraphics[width=0.6\columnwidth]{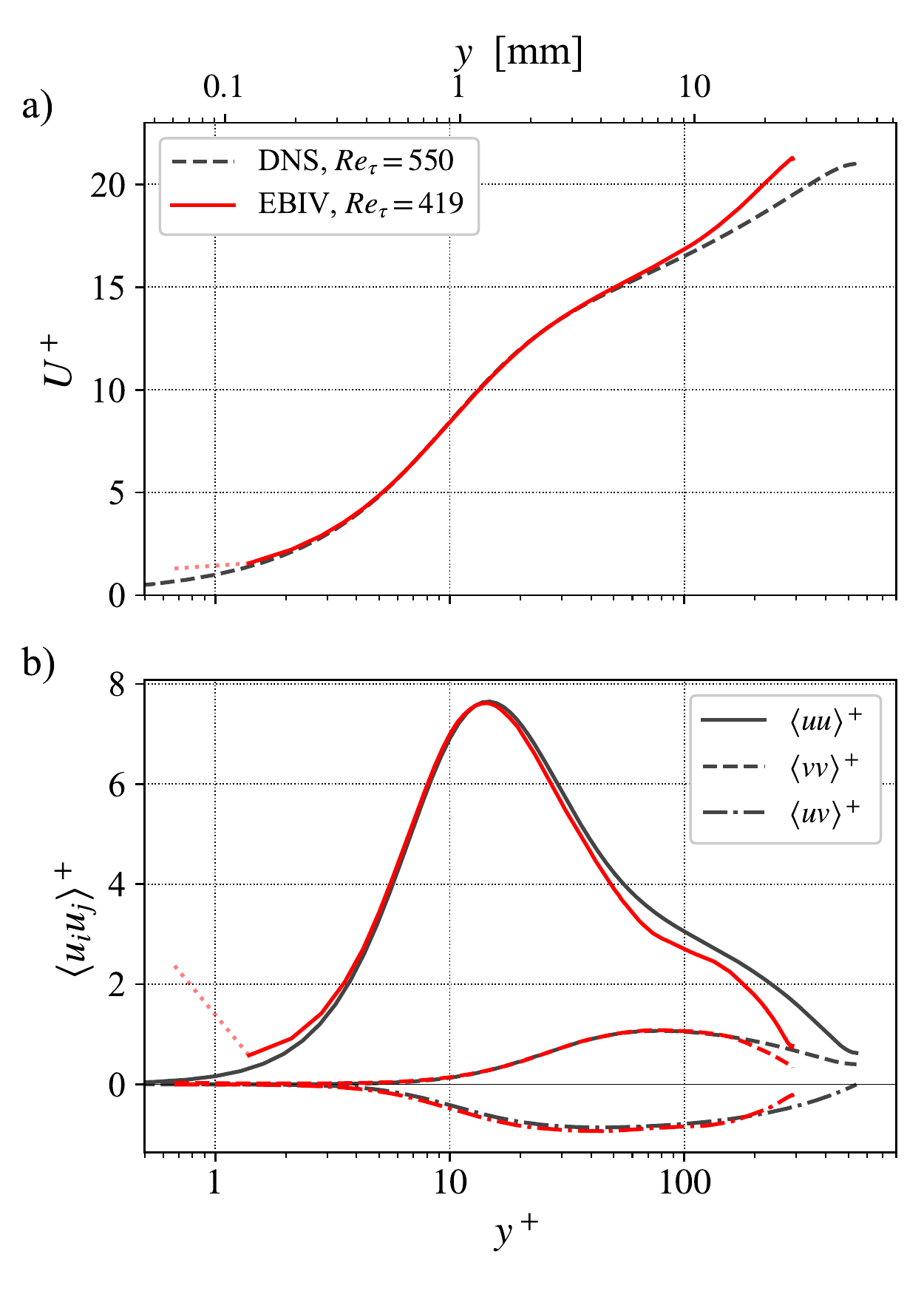}
\caption{Profiles of mean streamwise velocity $U$ (\textbf{a}) and  velocity variances $\langle u u \rangle$, $\langle v v \rangle$ and Reynolds stress $\langle u v \rangle$ (\textbf{b}) obtained by pulsed light EBIV in a turbulent duct flow. Quantities are normalized by inner variables, friction velocity $u_\tau$ and viscous unit $\nu/u_\tau$.
DNS data is from \protect\cite{LeeMoser:2015}.
} \label{fig:turbbl_profiles}
\end{SCfigure}

Profiles of the mean and higher order statistics computed from the temporal velocity profile are provided in Fig.~\ref{fig:turbbl_profiles}. 
Owing to the high spatial resolution of the near wall profile (Fig.~\ref{fig:turbbl_profiles}a)  the mean velocity gradient at the wall can be estimated from which the shearing velocity, $u_\tau\!=\!0.167$\,m/s, and viscous length scale, $\nu/u_\tau\!=\!91\,\upmu$m, can be derived \citep{Willert:2015}.
Profiles of the velocity variances, normalized by viscous scaling, are presented in Fig.~\ref{fig:turbbl_profiles}b together with DNS data for a fully developed channel flow at a similar friction Reynolds number \citep{LeeMoser:2015}.
While there is good agreement between DNS and the EBIV results, the fact that the flow in the square duct at $x/D \!=\! 20$ is not fully developed may explain the discrepancies of the streamwise variance $\langle u u \rangle$ at $y^+ \!>\! 50$.

\begin{SCfigure}[\sidecaptionrelwidth][htb]
\small
%\textbf{a)}\hspace{0.45\columnwidth}\textbf{b)}\\
\includegraphics[width=0.6\columnwidth]{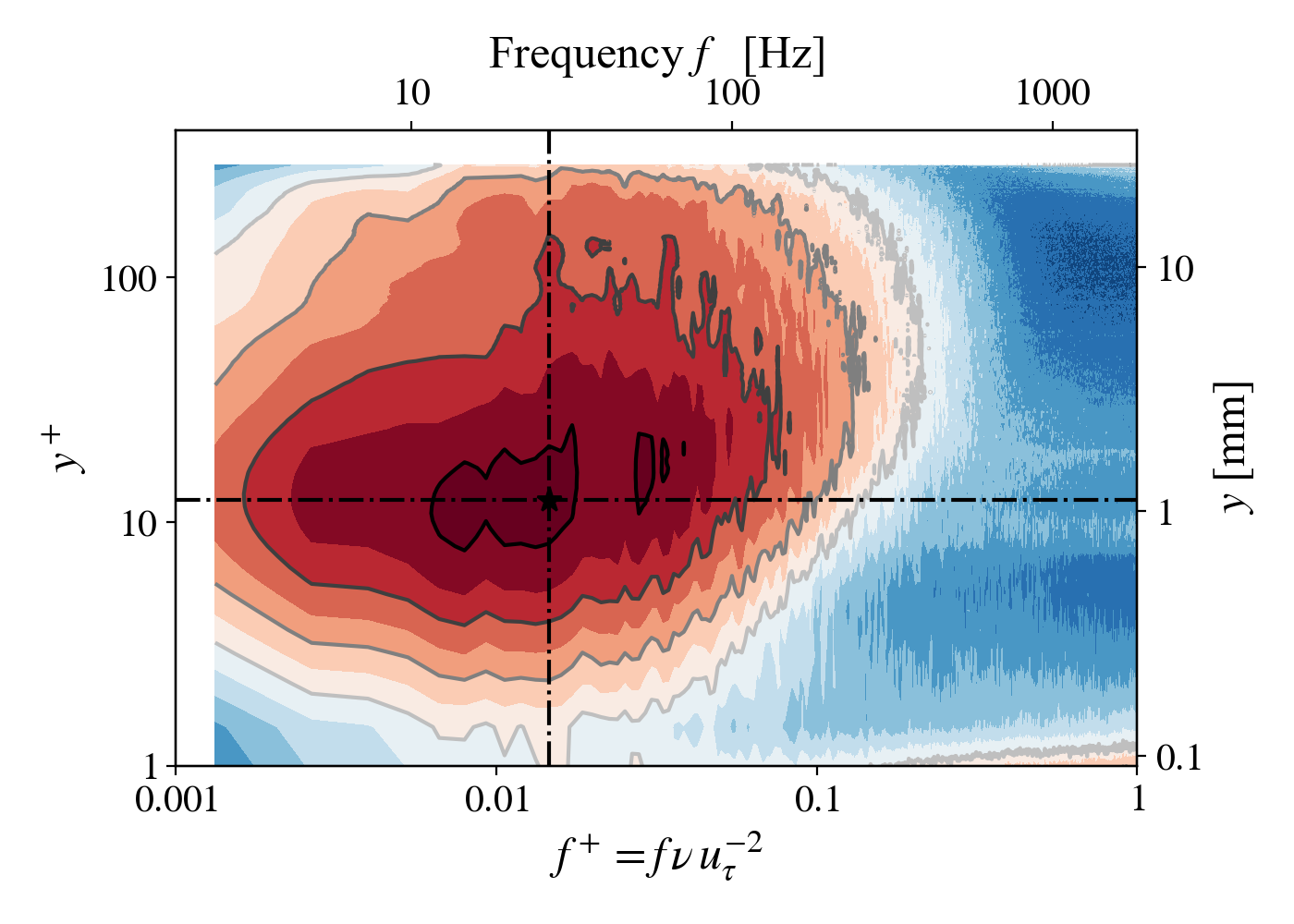}
\caption{Spatially resolved pre-multiplied power spectrum of the turbulent boundary layer within the square duct normalized by inner variables.  Dashed lines indicate position of energy maximum.}\label{fig:turbbl_spec2d}
\end{SCfigure}

Fig.~\ref{fig:turbbl_spec2d} further demonstrates the possibility of extracting spectral information from the time-resolved EBIV data. 
The spatially resolved spectrum of the streamwise velocity $u$ is compiled from pre-multiplied  power spectra of $u(y,t)$ for each wall-normal distance $y$.
The peak energy resides at a wall distance of $y^+\!=\!12.3$ (indicated by the horizontal dashed line), 
slightly below the maximum of the variance $\langle u u \rangle$ in Fig.~\ref{fig:turbbl_profiles}b where it is located at $y^+\!=\!14.4$.
The dominant frequency is 26.8\,Hz.

In the context of the present paper, Fig.~\ref{fig:turbbl_spec2d} acts as a placeholder of further analysis that can be performed with data obtained by the pulsed EBIV technique, such as space-time (2-point) correlations and perform spectral modal analysis (see e.g. \citealp{Willert:Lisbon:2022}). 

\subsection{Flow around a square rib}
\label{sec:rib-flow}
Structures such as ribs, pins, fins or other surface features are typically placed onto surfaces to enhance convective heat transfer. The internal cooling of turbine blades is a common application.
In the present case, a single square rib of size $H \!=\! 8.18$\,mm is placed spanwise in the square test section ($D \!=\! 76$\,mm) at $x \!=\! 1700$\,mm downstream of the channel entry.
The flow condition upstream of the rib is laminar with a free stream velocity of  $U_\infty \!\approx\! 2.0$\,m/s with a bulk Reynolds number of $\mathrm{Re}_b \!=\! 10\,000$ and 
a Reynolds number based on rib height of $\mathrm{Re}_H \!=\! 1\,100$. 

Image magnification is set at $m \!=\! 31.4$\,pixel/mm ($31.8\,\upmu$m/pixel).
The laser pulsing rate is fixed at 5\,kHz with pulse widths of $25\,\upmu$s ($130\,\upmu$J/pulse).

\begin{figure}[htb]
\small
\textbf{a)}\hspace{0.49\columnwidth}\textbf{b)}\\
\includegraphics[width=0.49\columnwidth]{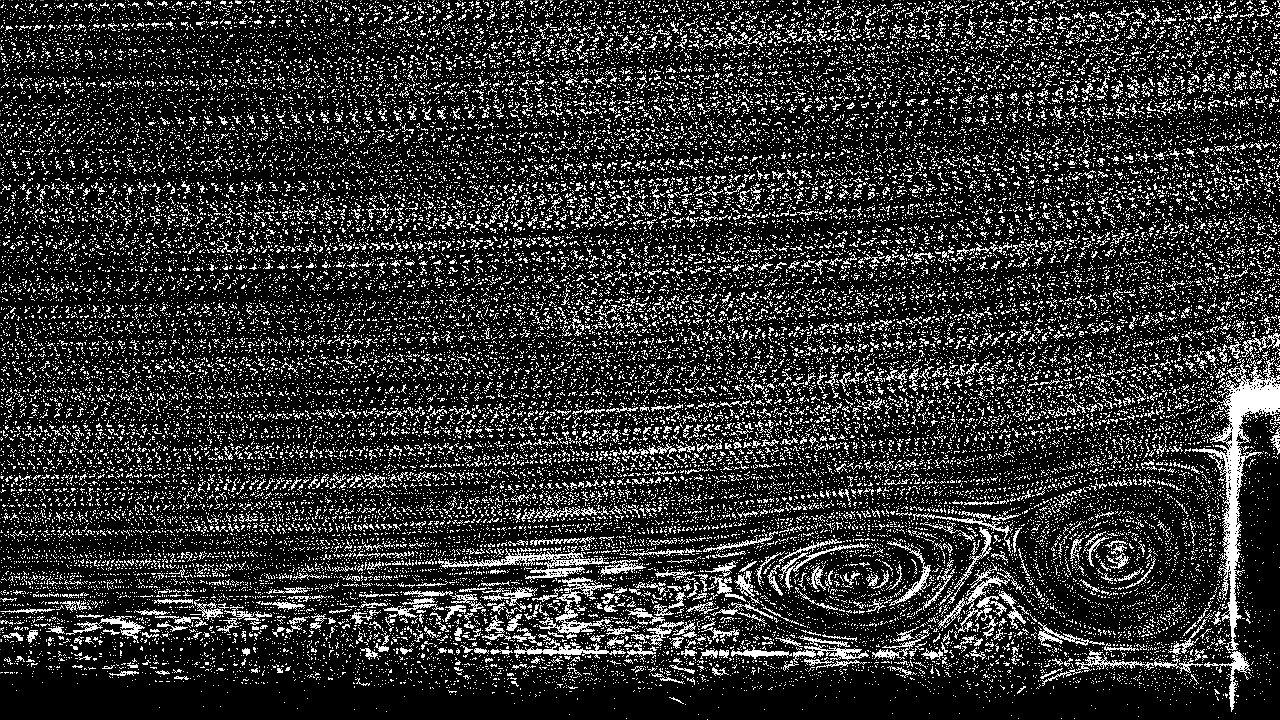}
\hfill
\includegraphics[width=0.49\columnwidth]{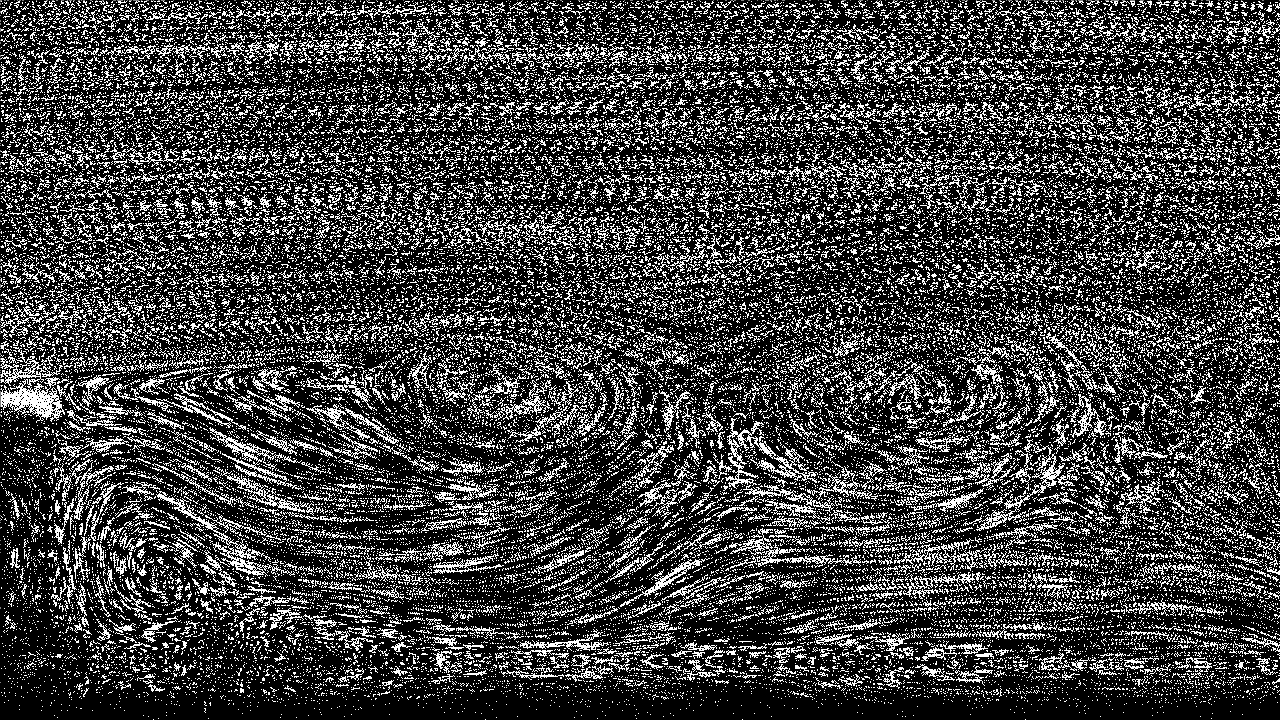}
\caption{Sample event data of the flow around a square rib of $H \!=\! 8.2$\,mm at $\textrm{Re}_H \!=\! 1\,100$ and pulsed illumination at $f_p \!=\! 5$\,kHz. With the  flow direction left-to-right, the time-slices capture 5\,ms of events or 25 laser pulses of laminar upstream region (\textbf{a}) and turbulent recirculation region in the immediate wake (\textbf{b}).
Animations of the raw event data are provided in the supplementary material.}
    \label{fig:rib-events}
\end{figure}

Event sequences of 10\,s duration are captured on a field of view of $41\!\times\!23\,\mathrm{mm}^2$ at different streamwise positions by traversing camera and light sheet.
Fig.~\ref{fig:rib-events} provides sampled event data for a duration of 10\,ms.
Event data rates vary between 42\,MEv/s and 50\,MEv/s with corresponding respective data rates of 111\,MB/s and 133\,MB/s.
With the light sheet coming from the top, the laser flare on the anodized aluminum surface of the square rib results in significant event generation within the otherwise dark region of the rib.
This could be considered an effect similar to the laser flare induced saturation on CCD (blooming).
Laser light passing through the glass of the channel results in less scattered light and mirror imaged particle event streaks can be observed.
Due to the pulsed illumination stationary particles and other dust like features on the wall also become visible.

Under laminar inflow conditions a series of nearly stable vortices (rollers) form upstream of the rib.
A shear layer is present above the rib, extends downstream and results in steady shedding of vortices about 3 rib heights downstream.
The immediate wake is characterized by a rather stable recirculation zone with velocities about one order of magnitude lower than the external flow.

Analoguous to the previously described boundary layer flow, the pseudo-image data of the flow around the square rib is processed with a multi-frame processing scheme ($N_f \!=\! 5$ frames) using samples of $40\! \times\!40$\,pixels ($1.3\!\times\!1.3\,\mathrm{mm}^2$).
With a seeding density estimated at $ppp \!=\! 0.0084$ about 13 particles are captured per sample on average (c.f. Table~\ref{tbl:ebiv_measdata}).

\begin{figure}[htb]
\small
%\textbf{a)}\hspace{0.49\columnwidth}\textbf{b)}\\
\includegraphics[width=\columnwidth]{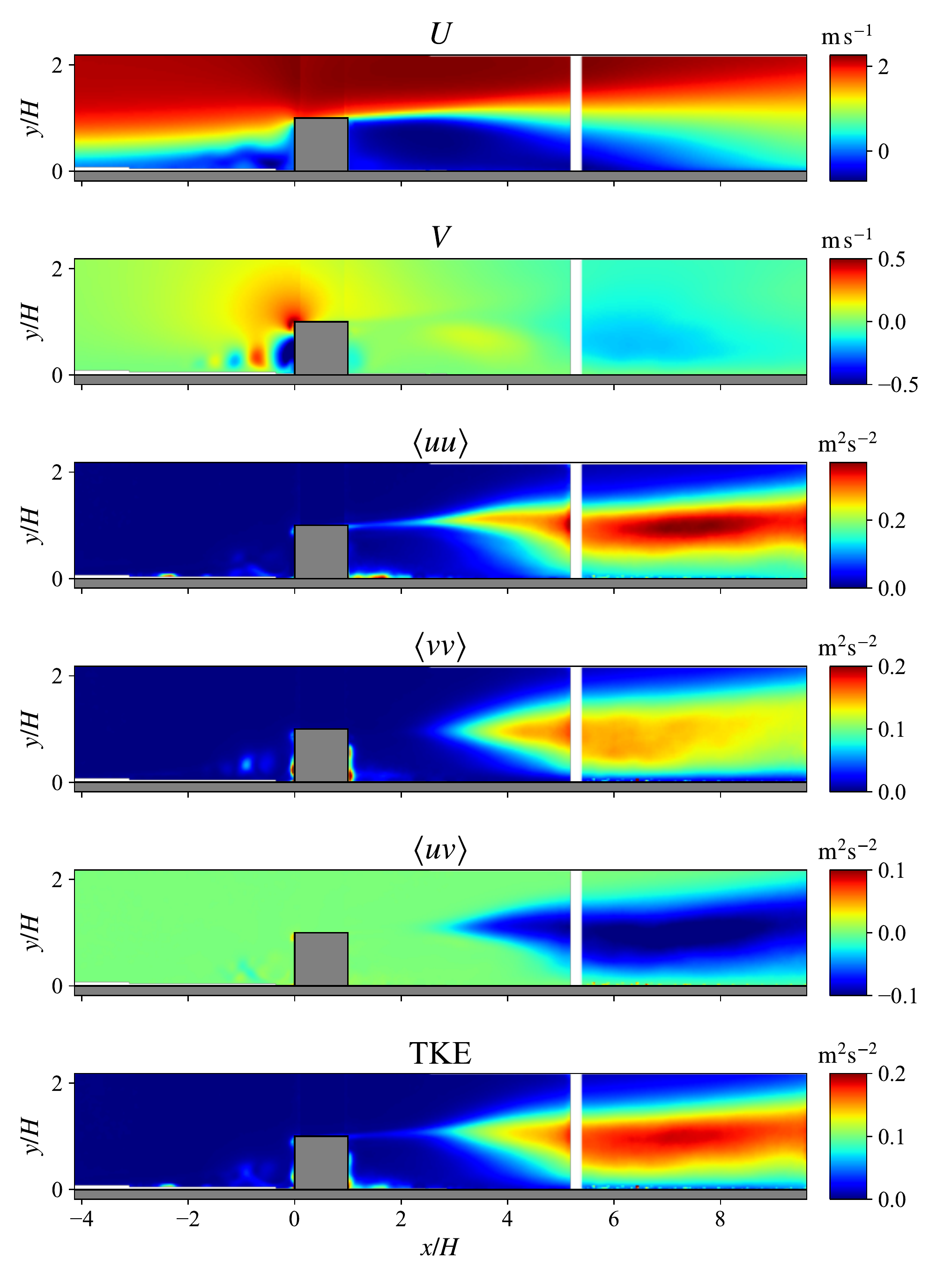}
\caption{Velocity statistics for the flow around a square rib with laminar inflow conditions obtained from 4 sampling domains of $36 \!\times\! 20\,\mathrm{mm}^2$ ($4.3\,H \!\times\! 2.4\,H$ ). Statistics are computed from $50\,000 \!\times 200 \upmu$s = 10\,s.}
    \label{fig:rib-stats}
\end{figure}

First and second order statistics of the flow field are plotted in Fig.~\ref{fig:rib-stats} compiled from 50\,000 (correlated) samples.

\begin{figure}[htb]
\small
\textbf{a)}\hspace{0.49\columnwidth}\textbf{b)}\\
\includegraphics[width=0.49\columnwidth]{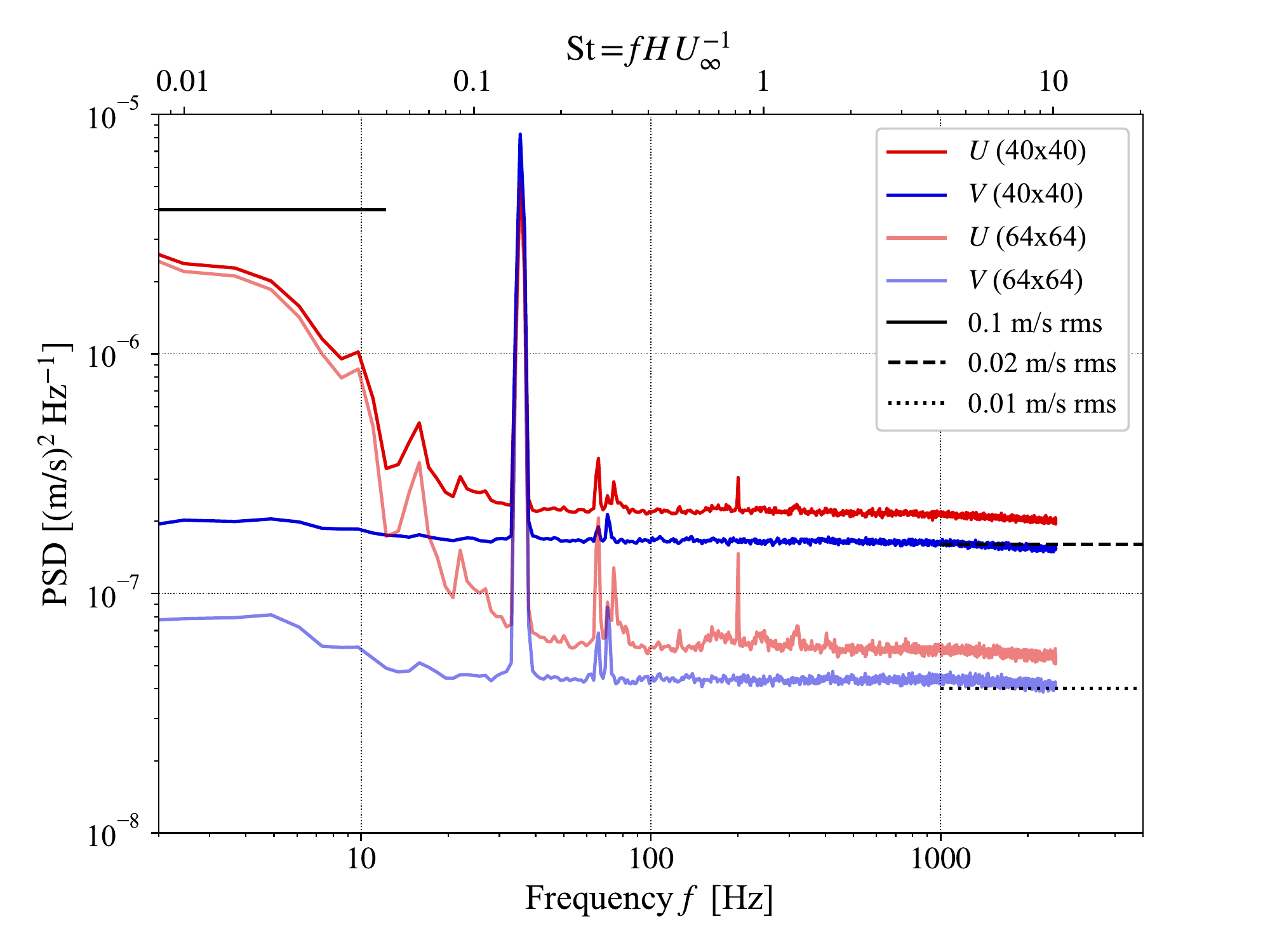}
\hfill
\includegraphics[width=0.49\columnwidth]{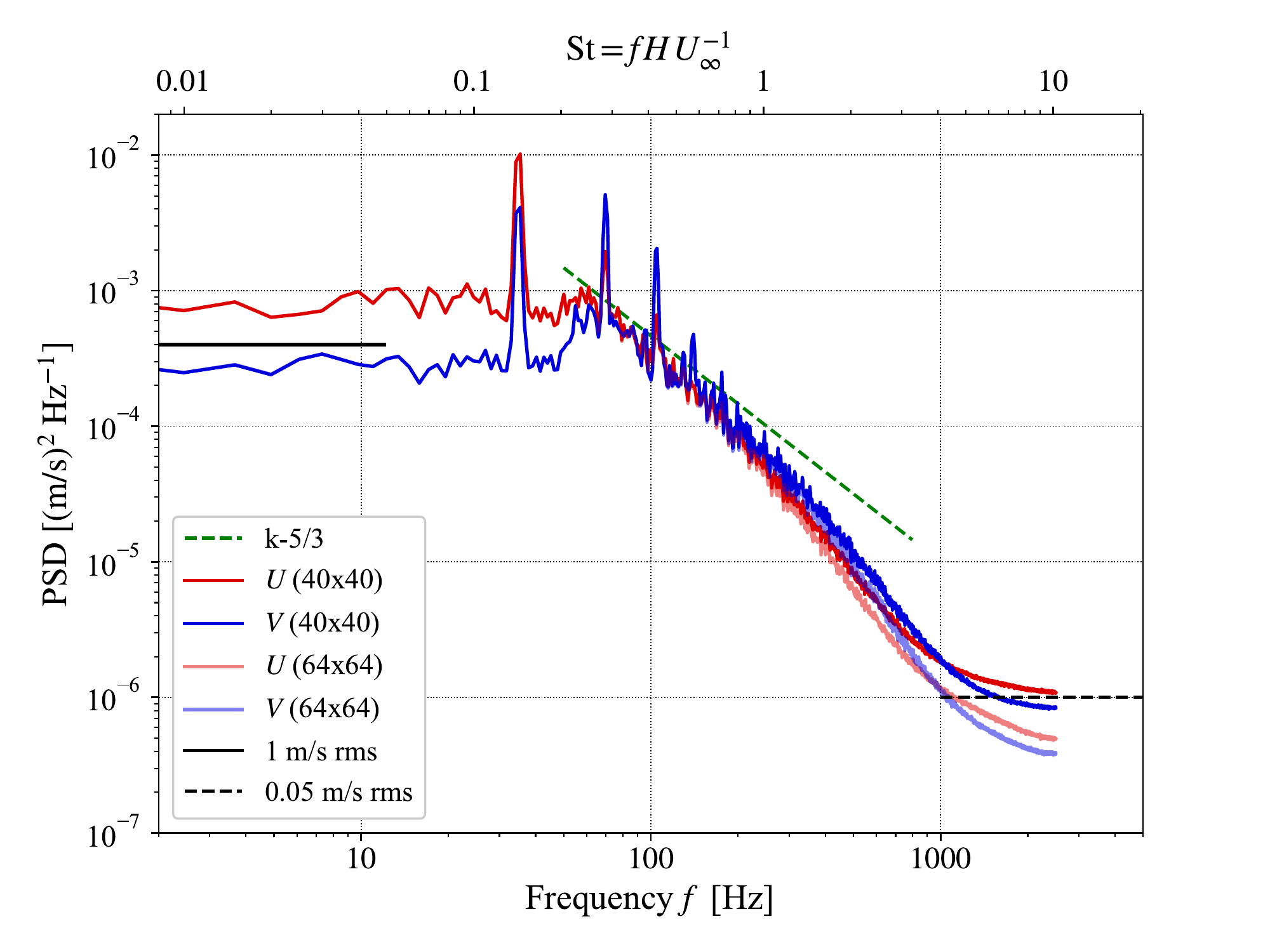}
\caption{Frequency spectra for both velocity components ($u,v$) of the flow around a square rib obtained in the laminar upstream region (\textbf{a}) and in the turbulent wake of the at $x/H = 8$ (\textbf{b}). 
} \label{fig:rib-spectra}
\end{figure}

Fig.~\ref{fig:rib-spectra} show spectra obtained both in the laminar region upstream of the rib and downstream in the turbulent wake.
The pronounced peak at $f \!=\! 35.4$\,Hz with a reduced frequency of $\mathrm{St} \!=\! f\,H\,U_\infty^{-1} \!=\! 0.145$ is the signature of the periodic vortex shedding in the shear layer between outer flow and wake region.
It is present for both velocity components and is also ``felt'' in the upstream region (Fig.~\ref{fig:rib-spectra}).
A low frequency modulation is present in the streamwise velocity component $u$ at frequencies below 10\,Hz and is due to a slight oscillation (pumping) of the mean flow in the channel. Its magnitude is about 3 orders of magnitude weaker than the spectral content in the turbulent wake and therefore cannot be observed in the spectra obtained in the wake flow (Fig.~\ref{fig:rib-spectra}b).

The frequency spectra permit an estimation of the noise level in the measurement. In the upstream laminar region (Fig.~\ref{fig:rib-spectra}a) the noise floor is constant across the frequency spectrum, aside from the signatures of the vortex shedding and the slight modulation of the streamwise velocity component.
The increase of the sampling window from $40\!\times\!40$\,pixel to $64\times\!64$\,pixel results in the reduction of the noise level by roughly a factor of two, which is a direct consequence of the increased number of events entering in the velocity estimation.
The spectrum obtained in the wake region (Fig.~\ref{fig:rib-spectra}b) resembles that of a classical turbulent energy cascade overlaid with the strong signature of the vortex shedding.
At a frequency of about 1\,kHz the spectrum exhibits a roll-up indicating the presence of a noise floor at about $10^{-6} (\textrm{m/s})^2 / \textrm{Hz}$ corresponding to a \ac{rms} velocity fluctuation of 0.05\,m/s.
The increase of sampling window size from $40\!\times\!40$ to $64\!\times\! 64$\,pixel reduces the noise floor and steepens the slope in the dissipation range of the energy cascade.
A maximum velocity of $U_\infty = 2$\,m/s and a noise level of 0.02\,m/s in the quiescent (laminar) flow suggests a \ac{DVR} of 100:1 for the present measurement.
This is in agreement with the uncertainty estimates obtained from the laminar boundary layer (c.f. Fig.~\ref{fig:lambl_profile}b,c).

\begin{table}
  \caption{Overview of performed pulsed EBIV measurements.} \label{tbl:ebiv_measdata}
  \centering
\begin{tabular}{lcccccc}
  \hline
   Flow & ROI & $f_p$ & $ppp^1$ & size$^1$, $d_i$ & $N_p$ & $N_p$ \\[1pt] %
   & [pixel] &  &  & [pixel] & $(32\times 32)$ & $(48\times 48)$ \\[2pt]
  \hline
  Turb. jet (water) & $1280\times 720$ & 2~kHz & 0.0059 & 1.46 & 6 & 8 \\
  & $1280\times 720$ & 4~kHz & 0.0033 & 1.39 & 3.4 & 8 \\
  & $1280\times 720$ & 5~kHz & 0.0034 & 1.37 & 3.5 & 14 \\[2pt]
  & $320\times 720$  & 2~kHz & 0.0153 & 1.49 & 16 & 35 \\
  & $320\times 720$  & 4~kHz & 0.0132 & 1.41 & 14 & 31 \\
  & $320\times 720$  & 5~kHz & 0.0128 & 1.39 & 13 & 30 \\
  & $320\times 720$  & 10~kHz & 0.0090 & 1.29 & 9 & 21 \\[2pt]
  & $1280\times 200$ & 5~kHz & 0.0111 & 1.42 & 11 & 26 \\
  & $1280\times 200$ & 10~kHz & 0.0078 & 1.37 & 8 & 18 \\[2pt]
  (PIV) & $1392\times 1040$ & 4~Hz & 0.0088 & 2.62 & 9 & 20 \\[2pt]
  \hline
  Lam. BL (air) & $320\times 720$  & 5~kHz & 0.0144 & 1.29 & 15 & 33 \\
  TBL (air) & $1280\times 320$ & 5~kHz & 0.0115 & 1.45 & 12 & 27 \\
  Square rib (air) & $1280\times 720$ & 5~kHz & 0.0084 & 1.25 & 9 & 19 \\[2pt]
\hline
\end{tabular}
\small{$^1$) Details on estimation of $ppp$ and particle size $d_i$ are provided in appendix section~\ref{sec:Appx:seeding density}.}
\end{table}

\section{Discussion}
\label{sec:discussion}
Due to bandwidth limitations of the sensor/camera the event rate that can be continuously acquired is limited.
%By impulsively illuminating the scene with pulsed light, the events are only generated by the particles within the scene.
In comparison to continuous illumination, where particles continuously generate events during their movement, the pulsed illumination results in spatially confined event generation (``frozen" particles).
Hence, the same number of particles will produce less events per unit time with pulsed illumination.
The previous experiments have shown that up to $40\!\cdot\! 10^6$\,events/s are feasible with pulsed particle illumination.
Above $40\!\cdot\!10^6$\,events/s the arbiters begin to saturate and can no longer process the pending events within the ``framing" rate imposed by the pulsing laser light.
In comparison to continuous illumination more particles can be captured; under ideal circumstances each particle would produce exactly one event per light pulse.
Particle streaking due to finite laser pulse duration, limited optical resolution (blurring) and non-uniform scattering intensity will result in particles producing multiple events per laser pulse.

Given that the number of reliably captured events per time-unit is constant, the number of particles per pseudo-frame reduces proportional to the increase of the laser pulsing rate which determines the effective frame rate of the pseudo-images.
For the present hardware limit of (about) $40\!\cdot\! 10^6$ events/s at best 8\,000 particles can be tracked simultaneously by the HD sensor at a laser pulsing rate of $f_p \!=\! 5$\,kHz, assuming that each particle produces one event per pulse on average. In practice it was found that each particle produced on average about 1.5 events in air and about 2.0 events in water, with these values being strongly influenced by imaging conditions (focus, lens quality, optical distortion, random event noise).
Increasing the pulse energy or the particle image density (\ac{ppp}) results in a saturation of the sensor's arbiters with consequential loss of timing information.
With the arbiters saturated, events can no longer be assigned to specific light pulses; the histogram of the event temporal distribution becomes uniform without distinct peaks corresponding to the light pulsing frequency (c.f. Fig~\ref{fig:waterjet-saturated}d).

Compared to the continuous illumination of the particles, the pulsed illumination results in a nearly constant particle image density, regardless of the flow velocity.
This is best explained by considering that for continuous illumination faster moving particles scatter less light per unit time onto a given pixel such that event rate reduces up to the point of producing no events.
With pulsed illumination the respective pixels receive the same amount of light (photons) regardless of the particle velocity, assuming sufficiently short pulses to prevent particle image streaking.
The finite pulse width of the low cost (engraving) lasers used herein results in a certain amount of image streaking for faster moving particles; that is, multiple spatially adjoining events are produced during the laser pulse.
While more costly, Q-switched high-speed lasers with narrow pulse widths in the 100\,ns range would further improve the quality of the acquired event data.

The pulsed light also makes stationary particles visible which would normally be hidden under constant illumination conditions. In previous implementations of EBIV the lack of events in quiescent regions of a flow have resulted in data loss \citep{Willert_EBIV:2022}.
If timing and pulse energies are carefully adjusted, the pulsed illumination does not result in an avalanche of events that would otherwise saturate the sensor. Background and surface scattering is generally less in comparison to PIV recording under similar illumination conditions.

\section{Conclusion}
By being able to clearly associate captured events from the EBV sensor with a given light pulse permits a more reliable tracking of particles because the timing uncertainty (i.e. latency)  introduced by the sensor's arbiters can be accounted for.
This also opens the possibility to extend the method for 3d tracking of particles using multiple synchronized cameras.
In previous 3d tracking implementations the timing uncertainty (i.e. latency and jitter) of the recorded events imposed significant uncertainties and required special filtering approaches in the 3d reconstruction of the space-time particle tracks (c.f. \citealp{Borer:2017}).

Within this work the event-data was analyzed using correlation-based algorithms that rely on pseudo-images rendered as an intermediate step. 
Alternatively, direct event-tracking schemes, derived e.g. from established \ac{PTV} algorithms, should be able to provide reliable Lagrangian tracking data of the imaged particles with considerably less computational effort. 
The track data could then be projected onto an Eulerian frame to obtain flow maps matching those produced by the correlation-based schemes presented herein \citep{Gesemann:2016,Godbersen:2020}.

The high sensitivity of the \ac{EBV} sensor reduces the demand for high-power laser illumination.
The low cost of the event-based cameras in comparison to high-speed cameras of similar frame rate (10\,kHz at 1\,MPixel) makes 4d Lagrangian particle tracking viable and affordable.
Also, the EBV camera hardware has a low power consumption (1--5\,W), is very compact (cube of 30--40\,mm, without lens) and has a low weight with the production versions of these cameras weighing about 50\,g. 
Nonetheless, it should be mentioned that the lack of grayscale information along with the bandwidth limitations of currently available event cameras results in a reduction of achievable particle image densities compared to imaging setups using conventional high-speed framing cameras.

%\begin{table}[h]
%\begin{center}
%\begin{minipage}{174pt}
%\caption{Caption text}\label{tab1}%
%\begin{tabular}{@{}llll@{}}
%\toprule
%Column 1 & Column 2  & Column 3 & Column 4\\
%\midrule
%row 1    & data 1   & data 2  & data 3  \\
%row 2    & data 4   & data 5\footnotemark[1]  & data 6  \\
%row 3    & data 7   & data 8  & data 9\footnotemark[2]  \\
%\botrule
%\end{tabular}
%\footnotetext{Source: This is an example of table footnote. This is an example of table footnote.}
%\footnotetext[1]{Example for a first table footnote. This is an example of table footnote.}
%\footnotetext[2]{Example for a second table footnote. This is an example of table footnote.}
%\end{minipage}
%\end{center}
%\end{table}

%\backmatter
%
%\bmhead{Acknowledgments}
%Acknowledgments are not compulsory. Where included they should be brief. Grant or contribution numbers may be acknowledged.

\appendix

\section{Declarations}

\begin{itemize}
%\item Funding
\item Conflict of interest:
There are no conflicts to declare.
%\item Ethics approval
%\item Consent to participate
%\item Consent for publication
%\item Availability of data and materials
%\item Code availability
%\item Authors' contributions
\end{itemize}

%\begin{appendices}
%

\section{Estimation of particle image density}\label{sec:Appx:seeding density}
Compared to conventional frame-based imaging the estimation of particle image density from asynchronous event data is slightly different due the fact that particles appear at distinct times. Therefore, a measure of particle image per pixel and time would be more appropriate.
In the case of pulsed illumination of particles, event times are assigned with the pulse timing as described in the preceding manuscript (c.f. Fig.~\ref{fig:pulsed_illumination}) allowing the event-data to be rendered as pseudo-images.
These pseudo-images can then be used in the estimation of particle image density in \textit{\ac{ppp}} allowing a comparison to frame-based imaging approaches.

Depending on scattering behavior and imaging conditions, a given particle illuminated by a light pulse can produce multiple events both spatially and temporally. Collected into pseudo-images this results in clusters of pixels representing individual particle images that can be counted. The cluster count is then used for the estimation of the particle image density in \textit{\ac{ppp}} and the particle image size $d_i$. 
Here, the mean particle image size $d_i$ is defined as the square root of the average cluster size.
Also included in the \textit{\ac{ppp}} estimate is the random event noise that varies depending on the EBV sensor's bias settings.
For the presented measurements this random noise is typically less than a percent but may reach up to several percent (see e.g. Fig.~\ref{fig:waterjet-saturated}c).
The random event contribution could be estimated by applying Lagrangian particle tracking which separates noise from particle image tracks.

For the PIV data presented in Sect.~\ref{sec:ebiv-waterjet} and Table~\ref{tbl:piv_ebiv} the particle image density is estimated by first subtracting the intensity mean of all images from an individual PIV recording.
The residual image is then binarized resulting in clusters that can be sized and counted as for the event-based pseudo-images.
Due to the mean intensity subtraction the actual \textit{\ac{ppp}} for the PIV recordings may actually be higher than the estimates.

\section{Supplementary material}
Animated sequences of the acquired event data  are provided as supplementary material.
All sequences were originally recorded with a laser pulsing frequency of $f_p \!=\! 5$\,kHz with pseudo-frames incremented by $1/f_p \!=\! 200\,\upmu$s. At a playback speed of 25 frame/s the events display at $0.005\times$ actual speed.
Duration of all sequences is 50\,ms.
Only positive events are rendered.

\begin{itemize}
	\item turbulent jet in water:  2\,ms time-slices (10 pulses)
	\item laminar boundary layer:  1\,ms time-slices (5 pulses)
	\item flow around rib (upstream location):  2\,ms time-slices (10 pulses)
	\item flow around rib (downstream location):  2\,ms time-slices (10 pulses)
\end{itemize}

%An appendix contains supplementary information that is not an essential part of the text itself but which may be helpful in providing a more comprehensive understanding of the research problem or it is information that is too cumbersome to be included in the body of the paper.
%
%\end{appendices}

%%===========================================================================================%%
%% If you are submitting to one of the Nature Portfolio journals, using the eJP submission   %%
%% system, please include the references within the manuscript file itself. You may do this  %%
%% by copying the reference list from your .bbl file, paste it into the main manuscript .tex %%
%% file, and delete the associated \verb+\bibliography+ commands.                            %%
%%===========================================================================================%%

%\bibliography{EBIV_lit_short}% common bib file
\bibliographystyle{unsrtnat}
\bibliography{EBIV_lit}% common bib file
%% if required, the content of .bbl file can be included here once bbl is generated
%%\input sn-article.bbl

\end{document}